%% file: arXiv_version.tex
\documentclass{IEEEtran}
\usepackage{cite}
\usepackage{graphicx}
\usepackage{amsmath}
\usepackage{array}
\usepackage{mdwmath}
\usepackage{amssymb}
\usepackage{mdwtab}
\usepackage{stfloats}
\usepackage[tight,footnotesize]{subfigure}
\usepackage{amsmath,amsthm}
\usepackage{threeparttable}
\usepackage{xcolor}
\usepackage{url}
\usepackage{algpseudocode}
\usepackage{multicol}
\usepackage{multirow}
\usepackage{setspace}
\usepackage{bm}
\usepackage{enumitem}
\algrenewcommand{\algorithmicrequire}{\textbf{Input:}}
\algrenewcommand{\algorithmicensure}{\textbf{Output:}}

\newtheorem{remark}{\it  Remark}

\newtheorem{definition}{\it  Definition}

\newcommand{\bcal}[1]{\boldsymbol{\mathcal{#1}}}
\newcommand{\hatbf}[1]{\hat{\mathbf{#1}}}

\newcommand{\comments}[1]{}
\definecolor{green1}{rgb}{0, 0.6, 0.3}  
\definecolor{blue1}{rgb}{0, 0.447, 0.741} 
\definecolor{red1}{rgb}{0.8, 0.15, 0}

\usepackage{hyperref}
\hypersetup{
    colorlinks=true,
    linkcolor=red,
    citecolor=red,
    urlcolor=blue,
    breaklinks=true}

\usepackage[nolist]{acronym}
\input{acronyms}

\begin{document}

\title{Spatial Bandwidth Asymptotic Analysis for 3D Large-Scale Antenna Array Communications} 
	\author{Liqin~Ding,~\IEEEmembership{Member,~IEEE,}
	    Jiliang~Zhang,~\IEEEmembership{Senior Member,~IEEE}
		and~Erik~G.~Str\"om,~\IEEEmembership{Fellow,~IEEE}
\thanks{This work was supported in part by  the European Union’s Horizon 2020 research and innovation programme under the Marie Skłodowska Curie Grant Agreement No. 887732 (H2020-MSCA-IF VoiiComm) and in part by the National Key R\&D Program of China under Grant
2021YFB3300900.}
\thanks{L. Ding and E. G. Str\"om are with the Department of Electrical Engineering, Chalmers University of Technology, Gothenburg, Sweden.}
\thanks{J. Zhang is with the College of Information Science and Engineering, Northeastern University, Shenyang, China.}}
	\markboth{ } 
	{Ding \MakeLowercase{\textit{et al.}}: Asymptotic Spatial Bandwidth Analysis for 3D Linear Large-Scale Antenna Arrays Communications}
	
\maketitle

\vspace{-3em}
\begin{abstract}
In this paper, we study the spatial bandwidth for line-of-sight (LOS) channels with linear large-scale antenna arrays (LSAAs) in 3D space. We provide approximations to the spatial bandwidth at the center of the receiving array, of the form $C R^{-B}$, where $R$ is the radial distance, and $C$ and $B$ are directional-dependent and piecewise constant in $R$. The approximations are valid in the entire radiative region, that is, for $R$ greater than a few wavelengths. When the length of the receiving array is small relative to $R$, the product of the array length and the spatial bandwidth provides an estimate of the available spatial degree-of-freedom (DOF) in the channel. In a case study, we apply these approximations to the evaluation of spatial multiplexing regions under random orientation conditions. The goodness-of-fit of the approximations is demonstrated and some interesting findings about the DOF performance of the channel under 3D and 2D orientation restrictions are obtained, e.g., that, under some conditions, it is better to constrain the receiving array orientation to be uniform over the unit circle in the 2D ground plane rather than uniform over the 3D unit sphere. 
\end{abstract}

\begin{IEEEkeywords}
Large-scale antenna array, degree-of-freedom, spatial bandwidth, spatial multiplexing.
\end{IEEEkeywords}

\section{Introduction}
\label{sec:1}

Research on wireless communication technologies utilizing \acp{LSAA} is experiencing significant growth and attention. Various schemes and concepts have been proposed, including \ac{LIS} \cite{hu2018beyond, dardari2020communicating}, \ac{ELAA} \cite{bjornson2019massive}, holographic \ac{MIMO} \cite{sanguinetti2022wavenumber, demir2022channel}, and extremely large-scale \ac{MIMO} (XL-MIMO) \cite{de2020non, lu2022communicating, wang2022extremely}, among others. The remarkably large physical size of \acp{LSAA} leads to unconventional advantages, such as precise beam focus at specific locations \cite{zhang2022beam} and spatial multiplexing under \ac{LOS} propagation conditions \cite{dardari2020communicating, yuan2021electromagnetic, sanguinetti2022wavenumber}, and brings wireless communication to a new regime where the information richness of the spatial domain has a more important role to play than ever before. To efficiently explore this regime, the need for a thorough understanding of the spatial characteristics of \ac{EM} waves cannot be overemphasized \cite{de2020non, lu2022communicating, bjornson2020power, demir2022channel, pizzo2022spatial}. 

One fundamental aspect that underlies several unconventional advantages, including the two mentioned above, is the presence of abundant spatial \acp{DOF} in \ac{LOS} channels with \ac{LSAA}. The related research can be traced back to the seminal work of O.~M.~Bucci and G.~Franceschetti in the late 1980s \cite{bucci1987spatial, bucci1989degrees}, where they studied the total number of spatial \acp{DOF} existing in the radiative \ac{EM} field radiated/scattered by sources confined within a fixed sphere. They revealed that the \textit{effective spatial bandwidth} of such EM field is limited and determined by the radius of the source sphere \cite{bucci1987spatial}, and that \textit{asymptotically} (in a sense that will become clear later), the total number of spatial \acp{DOF} is given by the number of Nyquist samples taken over an outer spherical observing domain \cite{bucci1989degrees}. The study was later generalized by Bucci et al.~to sources confined within a convex domain with rotational symmetry \cite{bucci1998representation}, where they introduced the concept of \textit{local spatial bandwidth} and demonstrated non-redundant Nyquist sampling by establishing a curve-linear coordinate system for the observing domain. Along this line, research has continued in the following two decades, notably by M.~Franceschetti and M.~D.~Migliore \cite{franceschetti2009capacity, franceschetti2011degrees, franceschetti2015landau, franceschetti2015information, franceschetti2018wave}. Most studies, and this paper too, consider the radiative part of a monochromatic EM field, which requires the observing domain to be at least a few wavelengths apart from the source region (the meaning will become clear in Section~\ref{sec:2})\footnote{When the observing region is close enough to capture the reactive component, additional \acp{DOF} are available \cite{franceschetti2015information}.}. 
Moreover, this paper studies the spatially-continuous setting, and, therefore, the obtained \ac{DOF} results serve as upper limits to what can be achieved by sampling the source and observing regions with small antennas\cite{hu2018beyond}.

It is important to note that the \ac{DOF} of a signal space with limited frequency support, observed over a finite duration, cannot be directly equated with the number of Nyquist samples taken for the reconstruction of a signal from this space \cite{slepian1976bandwidth, landau1980eigenvalue}. Their relation is rigorously addressed by Landau's eigenvalue theorem \cite{landau1980eigenvalue}. To be precise, for square-integrable time signals with frequency support limited to $[-B, B]$ Hertz, observed over a duration of $T$ seconds, the number of \acp{DOF} of the signal space is \cite[Eq.~(15)]{pizzo2022landau}
\begin{align}\label{eq:Landau}
\mathrm{DOF}_{\epsilon} = 2 BT + \frac{1}{\pi^2} \log\left( \frac{1-\epsilon}{\epsilon}\right) \log T + o(\log T).
\end{align}
where $0 < \epsilon <1 $ represents a tolerance level needed to rigorously define the dimensionality of the signal space (see \cite{slepian1976bandwidth} or \cite{franceschetti2015landau, pizzo2022landau}). In fact, $\mathrm{DOF}_{\epsilon}$ is equal to the number of eigenvalues (associated with a certain linear operator) that exceeds $\epsilon$. See Section~\ref{sec:2a} for more discussion related to the LOS channel. By plotting the eigenvalues (sorted in decreasing order), one finds that it begins with a region where the eigenvalues have similar strength, followed by a transition region centered at $2BT$ and with a width that grows as $\log T$, where the eigenvalues gradually decrease to be arbitrarily close to $0$ \cite{landau1980eigenvalue}. Hence, as $T \rightarrow \infty$, the width of the transit zone becomes negligible compared to $2BT$, regardless of the value of $\epsilon$. In this sense, we say that DOF is asymptotically given by $2BT$. We may also refer to $2B$ as the ``DOF density'' per unit of time in the same asymptotic sense. A completely symmetrical result can be obtained by fixing $T$ and increasing $B$; that is, we can think of $2T$ as the DOF density per unit frequency as $B$ becomes large. Landau's theorem extends to square-integrable, multidimensional, bandlimited signals and therefore to \ac{EM} fields \cite{franceschetti2015landau, pizzo2022landau}, which validates \cite{bucci1987spatial, bucci1989degrees, bucci1998representation} with a different approach\footnote{Bucci et al. addressed the little-o term by making the \textit{effective bandwidth} slightly larger than the spatial frequency support.}.

Unlike time signals whose frequency can be infinitely high, the support of spatial frequency of any radiative EM field is always confined within the interval $[-\frac{1}{\lambda}, \frac{1}{\lambda}]$ when observed over a 1D geometry\footnote{We choose to measure spatial frequency in cycles per meter, which corresponds directly to measure temporal frequency in Hertz. Moreover, the terms region/geometry/antenna/antenna array are used interchangeably in this paper.} \cite{hu2018beyond, Ding2022degrees}, and within a circular region of radius $\frac{1}{\lambda}$ when observed over a 2D or 3D geometry \cite{hu2018beyond, pizzo2022spatial, pizzo2022fourier, pizzo2022nyquist}, where $\lambda$ represents the wavelength. The maximum supports may come from an infinitely large source region or, equivalently, isotropic scattering. They lead to a ``spatial DOF density'' of $\frac{2}{\lambda}$ per meter or $\frac{\pi}{\lambda^2}$ per square meter, in the asymptotic sense (i.e., as the observation region goes to infinity or as $\lambda$ goes to $0$). In practical scenarios, when the operating frequency is given by design and the source and observation regions are bounded, the Nyquist sampling number can be quite small and the number of eigenvalues in the transition zone cannot be ignored. Despite this, the Nyquist sampling number still serves as a good measure of the number of significant \acp{DOF} available, as shown in \cite[Fig.~2]{pizzo2022landau} and our own study \cite[Fig.~8]{Ding2022degrees}\footnote{For capacity evaluation, an eigenmode analysis is needed to identify the strength of the eigenvalues in the transition zone. A rigorous formulation of the eigenproblem that applies to any bounded array geometry can be found in \cite{miller2019waves}. However, obtaining analytical expressions is generally unlikely, except for a few special cases (see \cite{ding2022shannon} for an example)}.

When the separation $R$ between the two regions is sufficiently large compared to their individual sizes, the paraxial approximation applies\cite{pizzo2022landau}, leading to the following well-known results: With two linear arrays, the 1D spatial frequency support has a width of approximately $\frac{L_s'}{\lambda R}$, leading to $\mathrm{DOF} \approx \frac{L_s' L_r'}{\lambda R}$; while with two planar arrays, the 2D spatial frequency support has an area of approximately $\frac{A_s'}{(\lambda R)^2}$, leading to $\mathrm{DOF} \approx \frac{A_s' A_r'}{(\lambda R)^2}$, where $L_s'$ ($A_s'$) and $L_r'$ ($A_r'$) represent the projected length (area) of the source and receiving arrays on a line/plane perpendicular to the direction of propagation\footnote{When the paraxial approximation holds, the propagation directions is sufficiently focused that any two points on the source and receiving arrays can be chosen to draw the connecting line and perform the orthogonal projection. }. Despite varying original settings, these results can be derived from multiple independent research works, including \cite{miller2000communicating, poon2005degrees}, and also through Bucci's approach, as discussed in \cite{Ding2022degrees}.

In the era of LSAAs, communication distances can be comparable to or even smaller than the size of the arrays, and variations in the geometric relationship between the arrays (e.g., distance, direction, orientation) have a significant impact on the DOF performance \cite{Ding2022degrees}. For very short distances, while a source LSAA can be treated as infinitely large, the orientation of the receiving array still needs to be taken into account\footnote{For the sake of discussion, we consider LSAA as the source array. Understandably, if they switch roles, the DOF of the LOS channel between them remains the same.}. Indeed, achieving the maximum frequency support of $\frac{2}{\lambda}$ or $\frac{\pi}{\lambda^2}$ (that is, $\mathrm{DOF} \approx \frac{2 L_r}{\lambda}$ or $\mathrm{DOF} \approx \frac{\pi A_r}{\lambda^2}$) is only possible when the receiving array is parallel to the LSAA. Moreover, the behavior of spatial bandwidth under intermediate distance conditions remains unclear. Although numerically evaluating spatial bandwidth and DOF using Bucci's approach is not difficult, it is still valuable to derive simple and interpretable expressions for spatial bandwidth in this region.

In this paper, we study the LOS channel between a linear LSAA and a linear receiving array, where the receiving array length is small relative to the communication distance. In this scenario, the DOF can be approximated as the product of the spatial bandwidth and the length of the receiving array. We obtain asymptotic expressions for spatial bandwidth $W$ in the form of $({L_s}/{R})^B$, where $A$ and $B$ are directional-dependent and piecewise constant in $R$. Hence, $\log(W)$ is asymptotically piecewise linear (affine) in $\log(R)$, where the slope is proportional to $-B$. We refer to this as a multi-slope model for the spatial bandwidth. These new asymptotic expressions are easily interpretable and provide insights and details beyond the results found in the literature. Our main contributions can be summarized as follows.

\begin{itemize}
    \item We derive asymptotic expressions for spatial bandwidth for the two orthogonal orientations that dominate the contribution in \ac{DOF}. The asymptotic expressions show distinct multi-slope linear decay relationships with the radial distance in the logarithmic domain. A good fit to the exact results is shown. 
    \item We obtain a simple dual-slope asymptotic expression for spatial bandwidth for the general orientation, at the cost of lower accuracy for radial distances below a few lengths of the source \ac{LSAA}. A location-and-orientation-dependent linear decay relationship with the radial distance is observed. 
    \item Based on the asymptotic expressions, we evaluate the DOF performance of the channel in a simple scenario under random orientation conditions, using two new concepts called the maximum and the expected spatial multiplexing regions. Different effects of 3D and 2D orientation constraints in the optimal and expectation senses are observed. 
\end{itemize}

We will first present the problem setting and preliminary results obtained in \cite{Ding2022degrees} in Section~\ref{sec:2}. Asymptotic results are given in Section~\ref{sec:3} and \ref{sec:4}, while the derivations are detailed in the appendices. The spatial multiplexing regions are studied in Section~\ref{sec:5}. Following the tradition of the signal processing community, column vectors are used in the paper. We adopt the notation $f(x) \sim \tilde{f}(x) \ \ (x \rightarrow x_0)$ for the asymptotically equivalent relation between $f(x)$ and $\tilde{f}(x)$, which means that $\lim_{x\rightarrow x_0} {f(x)}/{\tilde{f}(x)} =1$ \cite[Ch.~1.4]{de1958asymptotic}.

\section{Problem Setting and Preliminaries}
\label{sec:2}

\subsection{Problem setting and assumptions} 
\label{sec:2a}

Following \cite{Ding2022degrees}, we consider two continuous linear antenna arrays, $\mathcal L_s$ and $\mathcal L_r$, of lengths $L_s$ and $L_r$, consisting of infinitesimal Hertzian dipoles and study the wireless channel between them under the condition of \ac{LOS} propagation in 3D space.  We also assume a perfect and interference-free perception of the electric field using $\mathcal L_r$. Given the time-harmonic (with the $\exp(j\omega t)$ convention) current distribution over the source array, denoted by $\bcal{J}(\mathbf s)$, $\mathbf s \in \mathcal L_s$, the electric field perceived by the receiving array can be written as 
\begin{equation}\label{eq:1}
\bcal{E}(\mathbf{p}) = \int_{\mathcal{L}_s}  \bcal{G}(\mathbf{p}- \mathbf{s}) \bcal{J}(\mathbf{s}) \, \mathrm{d}\mathbf{s}, \quad  \mathbf p \in \mathcal L_r
\end{equation}
where the dyadic Green's function $\bcal{G}: \mathbb{R}^3\rightarrow \mathbb{C}^3$ is given by \cite[Appendix~I]{poon2005degrees}: 
\begin{align} 
\label{eq:2}
\bcal{G}(\mathbf{r}) =  
  &\frac{-j\omega\mu}{4 \pi r} \exp ( -j k_0 r ) \cdot \nonumber \\
  &\Big[  \underbrace{\left(  \mathbf{I}_3  - \hatbf{r} \hatbf{r}^{\mathrm{T}}  \right)}_{\text{``radiative''}}
		+\underbrace{ \Big( \frac{j}{k_0 r} -\frac{1}{k_0^2 r^2} \Big)   \left( \mathbf{I}_3 - 3 \hatbf{r} \hatbf{r}^{\mathrm{T}} \right) }_{\text{``non-radiative''}}
		\Big],
\end{align}
where $j = \sqrt{-1}$, $r \triangleq \| \mathbf r\|$ and $\hatbf r \triangleq \frac{\mathbf r}{r}$ representing the Euclidean norm and the unit vector of $\mathbf r$, $k_0 = \frac{2\pi}{\lambda}$ with $\lambda = \frac{2\pi c}{\omega}$ being the wavelength ($c$ is the propagation speed),  and $\mu$ is the permeability of free space, $\mathbf{I}_3$ is the $3\times 3$ identity matrix, and $(\cdot)^\mathrm{T}$ is the transpose operation. We require the minimum distance between $\mathcal L_s$ and $\mathcal L_r$ to be at least a few wavelengths ($\gtrsim 10 \lambda$) so that the non-radiative (i.e., reactive) term in \eqref{eq:2} can be neglected. Thereby, the \ac{LOS} channel can be interpreted as a linear deterministic mapping between two spatial signals $\bcal{J}(\mathbf s)$, $\mathbf s \in \mathcal L_s$, and $\bcal{E}(\mathbf{p})$, $\mathbf{p} \in \mathcal{L}_r$, by replacing $\bcal{G}$ in~\eqref{eq:2} by 
\begin{align}
    \bcal{G}_{\mathrm F}(\mathbf{r}) \triangleq \frac{-j\omega\mu}{4 \pi r} \exp ( -j k_0 r )\left(  \mathbf{I}_3  - \hatbf{r} \hatbf{r}^{\mathrm{T}}  \right).
\end{align} 

Using functional analysis terminology (see \cite{miller2000communicating, miller2019waves} and \cite[Chapter 3]{franceschetti2018wave}), the LOS channel acts as a compact linear operator between two Hilbert spaces, one for the source current distribution over $\mathcal L_s$, one for the electric field over $\mathcal L_r$.  The operator (denoted by $\mathsf{A}$) has a bandlimiting nature and the receiving signal space admits an orthonormal basis. The DOF of the signal space, that is, the dimension of the signal space, is the number of orthonormal basis functions needed to present any signal in this space to a required accuracy level $\epsilon$. The practical meaning of this accuracy level can be related to the noise level when we actually measure the electric field -- we are not able to identify those signals whose power is too far below the noise level. 
To obtain the DOF and an orthonormal basis, we need to obtain the singular value decomposition (SVD) of the operator $\mathsf{A}$, which can be done by solving the eigendecomposition problem of a Hermitian operator $\mathsf{A}^\dagger \mathsf{A}$ or $\mathsf{A} \mathsf{A}^\dagger$. $\mathrm{DOF}_\epsilon$ is the number of eigenvalues that exceed $\epsilon$. The results will also provide the optimal communication modes through this LOS channel \cite{miller2019waves, dardari2020communicating}. The above can be understood as the SVD of a discrete channel matrix $\mathbf{H}$, and its relationship with the eigendecomposition of $\mathbf{H}^\dagger \mathbf{H}$ or $\mathbf{H} \mathbf{H}^\dagger$, but in the continuous domain. 
Although the problem formulation is straightforward (see, e.g., \cite{miller2019waves}), obtaining analytical expressions is generally very difficult, except under a few special geometry conditions (see, e.g., \cite{ding2022shannon} for an example). Nevertheless, if one aims for a strict capacity evaluation of the LOS channel between two fixed-size arrays, the eigenmode analysis is necessary.

\begin{figure*}[!t]
	\centering \vspace{-1em}
	\includegraphics[width= .8\linewidth,trim= 0 0 0 0, clip]{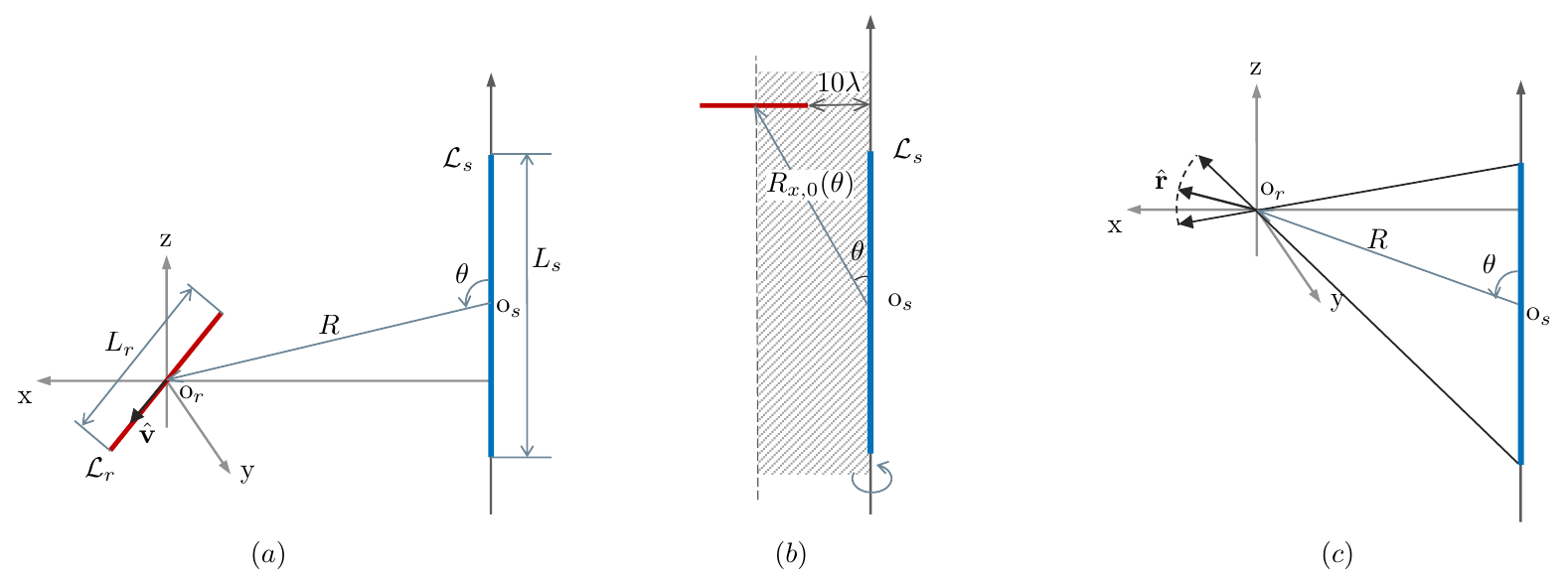} \vspace{-1em}
	\caption{Problem setting: (a) The local coordinate system and the parameters specifying the geometric relationship between the source-receiving array pair; (b) The minimum radial distance requirement \eqref{eq:R0}. (c) The range of $\hatbf{r}(0,q)$ is indicated by the dashed arc, which obviously depends not only on $L_s$, but also on $R$ and $\theta$. The actual range of spatial frequencies also depends on $\hatbf{v}$. 
 }  \vspace{-1em}
\label{fig1}
\end{figure*}

To unambiguously describe the geometric relationship between $\mathcal L_s$ and $\mathcal L_r$ in 3D space, in \cite{Ding2022degrees}, we proposed the use of $R$, the radial distance between the array centers $\mathrm{o}_s$ and $\mathrm{o}_r$, $\theta$, the polar angle defined from $\mathcal L_s$ to the $\mathrm{o}_s$-$\mathrm{o}_r$ connecting line, and $\hatbf v = (\hat{v}_{\mathrm x}, \hat{v}_{\mathrm y}, \hat{v}_{\mathrm z})^{\mathrm{T}}$, the unit directional vector of $\mathcal L_r$ in the \ac{LCS} defined at $\mathrm{o}_r$,  as depicted in Fig.~\ref{fig1}(a). The $\mathrm{z}$-axis of the \ac{LCS} is parallel to $\mathcal{L}_s$, the $\mathrm{x}$-axis lies on the $\mathrm{o}_r$-$\mathcal L_s$ plane and points away from $\mathcal L_s$, and the $\mathrm{y}$-axis is perpendicular to the $\mathrm{o}_r$-$\mathcal L_s$ plane. The unit vectors of the three axes are denoted by $\hatbf{e}_{\mathrm z}$, $\hatbf{e}_{\mathrm x}$, and $\hatbf{e}_{\mathrm y}$. It is the default coordinate system used in this paper unless otherwise noted. We restrict $\hat{v}_{\mathrm x} \geq 0$, which, since a sign change of $\hatbf{v}$ has no impact on the spatial bandwidth analysis,  does not limit generality. Moreover, $\theta \in (0,\pi)$ is assumed to prevent the two arrays from being colinear, and 
\begin{align}
\label{eq:R0}
    R \gtrsim \left( \frac{L_r}{2} +10\lambda \right)  \frac{1}{\sin\theta} 
\end{align}
ensures the applicability of $\bcal{G}_{\mathrm F}(\mathbf{r})$ for any orientation of $\mathcal L_r$ at the location under consideration. Note that the right-hand side of \eqref{eq:R0} is the minimum distance between $\mathcal L_r$ and $\mathcal L_s$ when $\mathcal L_r$ is orientated in the direction of the $\mathrm{x}$-axis, and that condition \eqref{eq:R0} restricts $\mathrm o_r$ to be outside the infinitely-long cylindrical region of radius $\frac{L_r}{2}+10\lambda$ centered around $\mathcal L_s$, as depicted in Fig.~\ref{fig1}(b). When $\mathcal L_r$ is orientated in the $\mathrm{z}$-axis direction, it suffices to require $R \gtrsim  10\lambda$. 

\subsection{Spatial bandwidth and K number}
\label{sec:2b}

Consider a source point $\mathbf s (q)$ on $\mathcal L_s$ with $q \in [- \frac{L_s}{2},  \frac{L_s}{2}]$, and a perception point $\mathbf{p}(l)$ on $\mathcal L_r$ with $l \in [-\frac{L_r}{2}, \frac{L_r}{2}]$. Let $\mathbf r(l,q) = \mathbf{p}(l) - \mathbf s (q)$ and denote its Euclidean norm and unit vector by $r(l,q)$ and $\hatbf{r}(l,q)$ respectively. We refer to $\mathbf{E}(l,q)  \triangleq \bcal{G}_\mathrm{F}(\mathbf{r}(l,q)) \bcal{J}(\mathbf{s}(q)) \, \mathrm{d}q$ as the wave component generated by the Hertzian dipole at $\mathbf s(q)$ in the electric field $\bcal{E}(\mathbf{p}(l))$. The \textit{spatial frequency} of $\mathbf{E}(l,q)$, denoted by $\kappa_{\hatbf{v}}(l, q)$, is defined as the directional derivative of the phase term $\frac{2\pi}{\lambda} r(l,q)$ along $\hatbf v$, divided by ${2\pi}$ \cite[Definition 1]{Ding2022degrees}. It can be easily verified that 
\begin{equation}
\label{eq:spatial_frequency}
\kappa_{\hatbf{v}}(l, q) 
	=  \frac{1}{\lambda}\,  \langle \hatbf{r}(l,q), \hatbf{v} \rangle 
 \quad \text{[cycle per meter]}, 
\end{equation} 
where $\langle \hatbf{r}, \hatbf{v} \rangle = \hatbf{r}^{\mathrm T} \hatbf{v} $. The range of $\kappa_{\hatbf{v}}(l, q)$ is thus bounded by $[-\frac{1}{\lambda}, \frac{1}{\lambda}]$. When $L_s$ is large enough, the change in $\hatbf{r}(l,q)$, and therefore in $\kappa_{\hatbf{v}}(l, q)$, can not be ignored for a fixed $l$ but different $q$. The \textit{local spatial bandwidth} of $\bcal{E}(\mathbf{p}(l))$, denoted by $w_{\hatbf{v}}(l; \mathbf{\Omega})$, where ${\mathbf{\Omega}}$ is a shorthand notation for the geometric parameters $(L_s, L_r, R, \theta)$, is defined as the difference between the maximum and minimum spatial frequencies of all the wave components in $\bcal{E}(\mathbf{p}(l))$. Namely, 
\begin{equation}
\label{eq:spatial_bandwidth_RCS}
	w_{\hatbf{v}}(l; \mathbf{\Omega}) \triangleq  \max_{ |q| \leq \frac{L_s}{2}} \, \kappa_{\hatbf{v}}(l, q) - \min_{|q| \leq \frac{L_s}{2}} \, \kappa_{\hatbf{v}}(l, q).
\end{equation} 
Note that \eqref{eq:spatial_bandwidth_RCS} differs from Bucci's original definition \cite{bucci1998representation} only by a factor of $2\pi$, which arises from our choice to measure spatial frequency in cycles per meter rather than radians per meter. Moreover, spatial bandwidth is ``double-sided'' in the sense of \eqref{eq:spatial_bandwidth_RCS}, in contrast to the convention that time signals with frequency support $[-B,B]$ is said to have bandwidth $B$.

Based on \eqref{eq:spatial_frequency} and \eqref{eq:spatial_bandwidth_RCS}, we can intuitively understand the spatial bandwidth geometrically, since $\kappa_{\hatbf{v}}(l, q)$ is determined by the projection of the unit vector $\hatbf{r}(l,q)$ onto another unit vector $\hatbf{v}$. In Fig.~\ref{fig1}(c), for an arbitrary location for the receiving array center $\mathrm{o}_r$,  we represent the range of $\hatbf{r}(0,q)$, i.e., $\{\hatbf{r}(0,q): q \in [- \frac{L_s}{2},  \frac{L_s}{2}] \}$, using a dashed arc of unit radius. Obviously, this range depends not only on $L_s$, but also on $R$ and $\theta$. For different $\hatbf{v}$, the values of $q$ that lead to the maximum and minimum projections are different, and thus the spatial bandwidth is also different. Therefore, it is conceivable that, for different $\theta$ and $\hatbf{v}$, the decay of $w_{\hatbf{v}}(0; \mathbf{\Omega})$ with $R$ will show different patterns. For example, although the largest range of $\hatbf{r}(0,q)$ occurs at $\theta = \frac{\pi}{2}$, two receive arrays oriented along $\hatbf{v} = \hatbf{e}_{\mathrm z}$ and $\hatbf{v} = \hatbf{e}_{\mathrm x}$ will have very different spatial bandwidth behaviors as $R$ increases, as our analysis will show later.

When $\mathcal{L}_r$ is long enough relative to $R$, for different locations on $\mathcal{L}_r$, the range of $\hatbf{r}(l,q)$, i.e., $\{\hatbf{r}(l,q): q \in [- \frac{L_s}{2},  \frac{L_s}{2}] \}$, will vary. Hence, in general, $w_{\hatbf{v}}(l; \mathbf{\Omega})$ varies with $l$. As a result, the number of samples given by non-redundant Nyquist sampling, which we will call the \textit{K number} in this paper, is given by\cite{franceschetti2011degrees}
\begin{equation}
\label{eq:spatial_DoF_RCS}
	K_{\hatbf{v}}({\mathbf{\Omega}} )  =  \int_{ -\frac{L_r}{2}}^{\frac{L_r}{2}} w_{\hatbf{v}} (l; \mathbf{\Omega}) \, \mathrm d l. 
\end{equation}
If $\mathcal{L}_r$ is short relative to $R$, it is conceivable that the range of $\hatbf{r}(l,q)$ will not vary much with $l$. In this case, the spatial bandwidth can be considered constant, denoted by  $W_{\hatbf{v}} (\mathbf{\Omega})$, and the K number is calculated by simple multiplication (the analogy with the $2BT$ formula is clear): 
\begin{equation}
\label{eq:7} 
	{K}_{\hatbf{v}}({\mathbf{\Omega}} ) \approx W_{\hatbf{v}} (\mathbf{\Omega}) L_r.
\end{equation}
This is the case when the paraxial approximation applies, as discussed in Section~\ref{sec:1}. We also demonstrated in \cite{Ding2022degrees} in a specific example with $L_r = 40\lambda$ and $L_s = 400\lambda$, that this constant spatial bandwidth approximation is applicable under mild conditions on $R$ for two orthogonal orientations: $\hatbf{e}_{\mathrm z}$ and $\hatbf{e}_{\mathrm x}$, see \cite[Fig.~7]{Ding2022degrees}. We already know from \cite{Ding2022degrees} that the spatial bandwidth along $\hatbf{e}_{\mathrm z}$ and $\hatbf{e}_{\mathrm x}$ are significantly larger than along $\hatbf{e}_{\mathrm y}$. Hence, we will ignore the contribution of the $\hatbf{e}_{\mathrm y}$ direction to the DOF\footnote{For very small $R$, the $\hatbf{e}_{\mathrm y}$ direction can also contribute in DOF, albeit much less than the other two directions. To actually compute this contribution, the change in the $w_{\mathrm y}(l; \mathbf{\Omega})$ cannot be ignored. In fact, \cite[Eq. (44)]{Ding2022degrees} shows that the minimum value of $w_{\mathrm y}(l; \mathbf{\Omega})$ appears at the array center and is always $0$. On the other hand, $W_{\mathrm y}(\mathbf{\Omega}) = w_{\mathrm y}(0; \mathbf{\Omega}) =0$ may also be regarded as a valid approximation, considering the relative small K number $w_{\mathrm y}(l; \mathbf{\Omega})$ actually lead to.}.

\begin{figure*}[b]
\vspace{-.5em}
\hrule 
\small 
\begin{align} 
\label{eq:wz0}
w_{\mathrm z}(l; \mathbf{\Omega})  
&= \frac{1}{\lambda} \bigg(  \frac{l+ R\cos\theta + \frac{L_s}{2}}{\sqrt{(l\!+\! R\cos\theta\! +\! \frac{L_s}{2})^2 \!+\! (R\sin\theta)^2}} 
	-  \frac{l+ R\cos\theta - \frac{L_s}{2} }{\sqrt{(l\!+\! R\cos\theta \!-\! \frac{L_s}{2})^2 \!+\! (R \sin\theta)^2}} \bigg), 
\end{align}
\begin{align}
w_{\mathrm x}(l; \mathbf{\Omega})  
	&= \begin{cases} 
	     w_{\mathrm x1}(l; \mathbf{\Omega})  = \frac{1}{\lambda} \big( 1 - \frac{l+ R\sin\theta }{\sqrt{(l + R\sin\theta )^2 + (R |\cos\theta|  + \frac{L_s}{2} )^2}} \big), &  R  |\cos\theta| \leq \frac{L_s}{2}, \\
	    w_{\mathrm x2}(l; \mathbf{\Omega}) =  \frac{1}{\lambda} \big( \frac{l + R\sin\theta }{\sqrt{(l + R\sin\theta  )^2 + (R |\cos\theta|  - \frac{L_s}{2})^2}} 
	     	- \frac{l + R\sin\theta }{\sqrt{(l + R\sin\theta )^2 + (R|\cos\theta| + \frac{L_s}{2})^2}} \big),  &  R |\cos\theta| >  \frac{L_s}{2}.
	     \end{cases}  \label{eq:wx0} 
\end{align}
\end{figure*}

When the constant spatial bandwidth approximation holds, the local spatial bandwidth at any position $l$ can essentially be used as $W_{\hatbf{v}} (\mathbf{\Omega})$. A convenient choice is $w_{\hatbf{v}} (0; \mathbf{\Omega})$. The exact closed-form expressions of $w_{\hatbf{v}}(l; \mathbf{\Omega})$ for $\hatbf{v} = \hatbf{e}_{\mathrm z}$ and $\hatbf{v} = \hatbf{e}_{\mathrm x}$ derived in \cite{Ding2022degrees} are given in \eqref{eq:wz0} and \eqref{eq:wx0} {at the bottom of next page}. For convenience, we denote them by $w_{\mathrm z}(l; \mathbf{\Omega})$ and $w_{\mathrm x}(l; \mathbf{\Omega})$. For a general orientation $\hatbf{v}$, the exact derivation of $w_{\hatbf{v}}(l; \mathbf{\Omega})$  is challenging, even just for  $l=0$. We will therefore first perform asymptotic analysis to $w_{\mathrm z}(0; \mathbf{\Omega})$ and $w_{\mathrm x}(0; \mathbf{\Omega})$, and then formulate an asymptotic expression for $w_{\hatbf{v}}(0; \mathbf{\Omega})$ that is valid for arbitrary $\hatbf{v}$.

\section{Asymptotic Functions for \texorpdfstring{$\mathrm{z}$}{z} and  \texorpdfstring{$\mathrm{x}$}{x} Directions}
\label{sec:3}

We define $W_{\mathrm z}(R; \theta) = w_{\mathrm z}(0;\mathbf{\Omega})$ and $W_{\mathrm x}(R; \theta) = w_{\mathrm x}(0;\mathbf{\Omega})$ as functions of $R$ with a parameter $\theta$. We sometimes omit the argument $(R;\theta)$ for brevity. By plotting $\log(W_{\mathrm z})$ and $\log(W_{\mathrm x})$ against $\log(R)$, we observe $\theta$-dependent linear relationships for different ranges of $R$. This observation motivates the asymptotic analysis detailed in Appendices~\ref{Sec:Appendix1} and \ref{Sec:Appendix2}, and leads to two asymptotic functions $\tilde{W}_{\mathrm z}(R; \theta)$ and $\Tilde{W}_{\mathrm x}(R; \theta)$ summarized in Table~\ref{table1} and Table~\ref{table2}. These functions are formed by three or two asymptote segments, in the form of power-law functions ($k=1$, $2$, or $3$): 
\begin{equation}
\label{eq:asymptote_general}
    \Tilde{W}^{(k)}(R;\theta) =  A(\theta) \cdot \left(\frac{L_s}{R}\right)^{B(\theta)},
\end{equation}
that are valid in different $R$ regimes. We refer to $B(\theta)$ as the spatial bandwidth exponent (SBE) in this paper. Necessary explanations will be given in the first two subsections, followed by an examination of the goodness-of-fit. For the convenience of discussion, we will focus on the range $\theta \in (0, \frac{\pi}{2}]$, since $W_{\mathrm z}(R;\theta)$ and $W_{\mathrm x}(R;\theta)$ are both symmetric about $\theta = \frac{\pi}{2}$. 

\begin{table*}[!t]
\caption{Asymptote segments and critical distances for the multi-slope asymptotic function $\Tilde{W}_{\mathrm z}(R;\theta)$, valid for $R \gtrsim  10\lambda$; with $\eta (\theta) = \frac{\sin \theta}{ \sqrt{1 + 3 \cos^2 \theta}}$ in the expressions.}
\label{table1}
\centering 
\small
\renewcommand{\arraystretch}{1.5}
\begin{tabular}{|l l|} \hline
\hspace{-1em}
\begin{tabular}{p{.07\linewidth}|p{.38\linewidth}|}
\textbf{Segment} & \textbf{Expression} \\
\hline
 1 & $\tilde{W}_{\mathrm z}^{(1)}(R;\theta) = \frac{2}{\lambda}$ \\
 2 & $\tilde{W}_{\mathrm z}^{(2)}(R;\theta) = \frac{1}{\lambda} \cdot \frac{\sqrt{1- \eta^2(\theta)}}{(2\, |\cos\theta|\, )^{B_{\mathrm z,2}(\theta)}} \cdot \left(\frac{L_s}{R}\right)^{B_{\mathrm z,2}(\theta)}$ \\
 & where $B_{\mathrm z,2}(\theta) = \frac{ 1}{2} \left[\eta^2(\theta) + \eta^{-1} (\theta) \right]$ \\
 3 & $\tilde{W}_{\mathrm z}^{(3)}(R;\theta) = \frac{1}{\lambda} \cdot \sin^2\theta \cdot \frac{L_s}{R}$ \\ [2pt]
\end{tabular} 
& \hspace{-1.4em}
\begin{tabular}{|p{.45\linewidth}} 
\textbf{Critical distances} \\ \hline 
$R_{\mathrm z,1,2}(\theta) = \frac{L_s}{2\, |\cos\theta|} \Big(\frac{\sqrt{1- \eta^2(\theta)} }{2 } \Big)^{ \frac{ 1 }{B_{\mathrm z,2}(\theta) }}$ \\ [6pt]
$R_{\mathrm z,2,3}(\theta) =  \frac{L_s}{ 2\,|\cos\theta|} \left( \frac{\sqrt{1- \eta^2(\theta)}}{2 \sin^2\theta |\cos\theta|} \right)^{ \frac{ 1 }{B_{\mathrm z,2} (\theta)-1}} $ \\[6pt]
$R_{\mathrm z,1,3}(\theta) = \frac{1}{2} L_s \sin^2\theta $  \\
\end{tabular} \\ \hline 
\multicolumn{2}{| p{.97\linewidth}|}{\textbf{Formation rule for $\Tilde{W}_{\mathrm z}(R;\theta)$:} For $\theta \in (0, 0.3197\pi) \cup (0.3285\pi, 0.5\pi) \cup (0.5\pi, 0.6715\pi) \cup (0.6803\pi, \pi )$, use segments 1, 2, 3 with critical distances $R_{\mathrm z,1,2}(\theta)$ and $R_{\mathrm z,2,3}(\theta)$; for 
  $\theta \in [0.3197\pi, 0.3285\pi] \cup \{ 0.5\pi \} \cup [0.6715\pi, 0.6803\pi]$, 
  use segments 1, 3 with critical distance $R_{\mathrm z,1,3}(\theta)$.  } \\ \hline 
\end{tabular}\vspace{-1em}
\end{table*}

\begin{table*}[!t]
\caption{Asymptote segments and critical distances for the multi-slope asymptotic function  $\Tilde{W}_{\mathrm x}(R;\theta)$, 
valid for $R \gtrsim  \big( \frac{L_r}{2} +10\lambda \big)  \frac{1}{\sin\theta}$, with $\eta (\theta) = \frac{\sin \theta}{ \sqrt{1 + 3 \cos^2 \theta } }$ in the expressions. }
\label{table2} 
\centering \vspace{-.5em}
\small 
\renewcommand{\arraystretch}{1.6}
\begin{tabular}{|l l|} \hline
\hspace{-1em}
\begin{tabular}{p{.07\linewidth}|p{.38\linewidth}|} 
\textbf{Segment} & \textbf{Expression}  \\
\hline 
1 & $\tilde{W}_{\mathrm x}^{(1)}(R;\theta) = \frac{2}{\lambda}$ \\
 2 & $\tilde{W}_{\mathrm x}^{(2)}(R;\theta) = \frac{1}{\lambda} \cdot \frac{1 - \eta(\theta)}{(2 \,|\cos\theta |\, )^{B_{\mathrm x,2}(\theta) }} \cdot \left(\frac{L_s}{R}\right)^{B_{\mathrm x,2}(\theta)}$ \\
 & where $B_{\mathrm x,2}(\theta)  = \frac{1}{2} \left[ \eta^2(\theta) + \eta(\theta) \right]$ \\ 
 3 & $\tilde{W}_{\mathrm x}^{(3)}(R;\theta) = \frac{1}{\lambda} \cdot \sin^2\theta \cdot \frac{L_s}{R}$ \\
 3$^*$ &$\tilde{W}_{\mathrm x}^{(3^*)}(R;\theta) = \frac{1}{8\lambda} \cdot \left(\frac{L_s}{R}\right)^2$ \\
\end{tabular} 
& \hspace{-1.4em}
\begin{tabular}{|p{.45\linewidth}} 
\textbf{Critical distances} \\ \hline
{$R_{\mathrm x,1,2}(\theta) = \frac{L_s}{ 2\, |\cos\theta| } (1 - \eta(\theta))^{\frac{ 1 }{B_{\mathrm x,2}(\theta) }}$} \\ [5pt] 
{$R_{\mathrm x,2,3}(\theta) = \frac{L_s}{ 2\, |\cos\theta| } \left( \frac{ 1 - \eta(\theta) }{ 2 \sin\theta \cos^2\theta } \right)^{\frac{ 1 }{B_{\mathrm x,2} (\theta) - 1 }}$} \\ [5pt]
{$R_{\mathrm x,1,3}(\theta) = L_s \sin\theta\, |\cos\theta|$} \\ [5pt]
{$R_{\mathrm x,1,3^*}(\theta) = \frac{L_s}{\sqrt{8}}$ } \\
\end{tabular} \\ \hline
\multicolumn{2}{|p{.97\linewidth}|}{\small \textbf{Formation rule for $\Tilde{W}_{\mathrm x}(R;\theta)$:} For $\theta \in (0, 0.0225\pi] \cup [0.9775\pi, \pi )$, use segments 1 and 3 with critical distance $R_{\mathrm x,1,3}(\theta)$; for $\theta \in (0.0225\pi, \frac{\pi}{2})\cup (0.5\pi, 0.9775\pi)$, use segments 1, 2, and 3 with critical distances $R_{\mathrm x,1,2}(\theta)$ and $R_{\mathrm x,2,3}(\theta)$; for  $\theta = \frac{\pi}{2}$, use segments 1 and 3$^*$ with critical distance $R_{\mathrm x,1,3^*}(\theta)$.}\\
\hline
\end{tabular}\vspace{-1.5em}
\end{table*}

\subsection{Asymptotic expressions for \texorpdfstring{$\tilde W_{\mathrm z}(R;\theta)$}{z}}
\label{sec:3a}

\begin{figure}[!t]
\centering  
\subfigure[$B_{\mathrm z,2}(\theta)$]{ \includegraphics [width= .88\linewidth,trim= 30 280 30 290, clip]{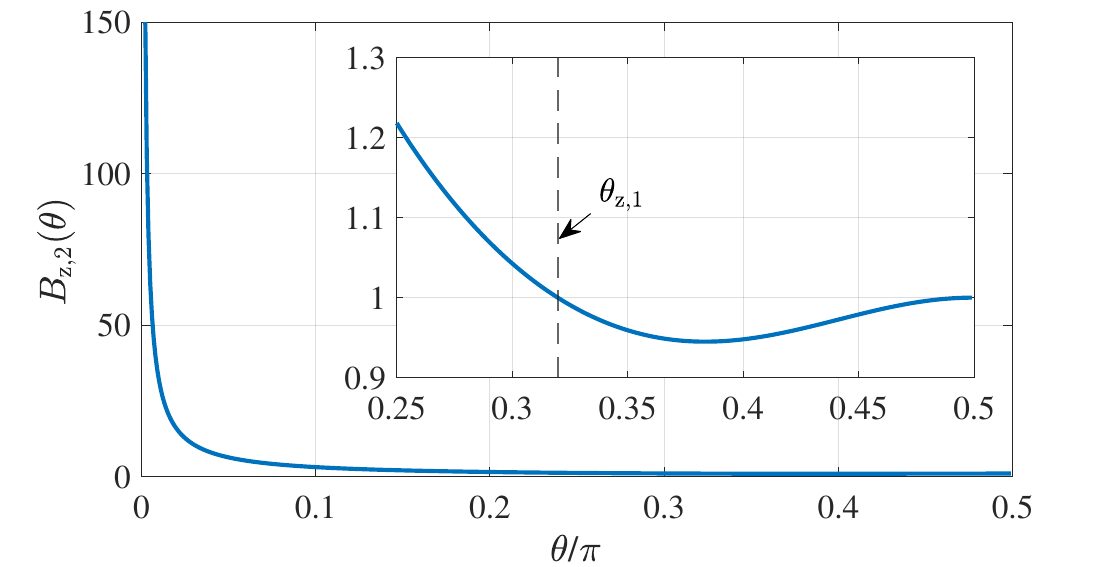}} 
\subfigure[Critical distances]{\includegraphics [width= .88\linewidth,trim= 30 280 30 290, clip]{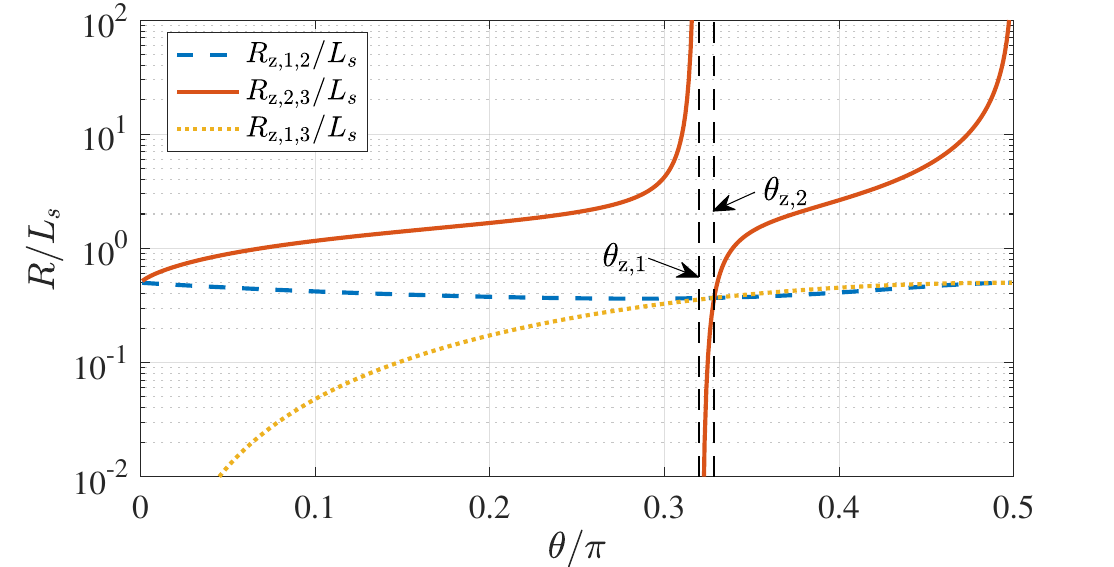}}\vspace{-0.5em}
\caption{SBE $B_{\mathrm z,2}(\theta)$ and critical distances (divided by $L_s$) for $\Tilde{W}_{\mathrm z}(R;\theta)$; with $ \theta_{\mathrm z,1} \approx 0.3197\pi$ and  $\theta_{\mathrm z,2} \approx 0.3285\pi$.} 
 \vspace{-1em}
\label{fig_parametersZ}
\end{figure}

The asymptotic analysis in Appendix~\ref{Sec:Appendix1} results in three asymptotes respectively for the small, medium, and large $R$ regimes, which are referred to as segments 1, 2, and 3 in Table~\ref{table1}. 
The SBEs of segments 1 and 3 are constant and given by $B_{\mathrm z,1} = 0$ and $B_{\mathrm z,3} =1$, while the SBE of segment 2 is $\theta$ dependent and given by
\begin{align}\label{eq:alpha}
    B_{\mathrm z,2}(\theta) = \frac{ 1}{2} \left[\eta^2(\theta) + {\eta ^{-1}(\theta)} \right] , 
\end{align}
where 
\begin{align}\label{eq:eta_1}
    \eta (\theta) = \frac{\sin \theta}{ \sqrt{1 + 3 \cos^2 \theta } }.
\end{align}
The curve of $B_{\mathrm z,2}(\theta)$ is plotted in Fig.~\ref{fig_parametersZ}~(a) for $\theta \in (0, \frac{\pi}{2}]$. It can be seen that as $\theta$ increases, $B_{\mathrm z,2}(\theta)$ drops rapidly in the beginning and to below $1$ after a critical value $\theta_{\mathrm z,1}$, but eventually increases to $1$ again at $\theta =\frac{\pi}{2}$. By solving equation $B_{\mathrm z,2}(\theta) = 1$, it is found that $ \theta_{\mathrm z,1} = \arccos \left( \sqrt{ \frac{1}{2\sqrt{5} -1} } \,\right) \approx 0.3197\pi$.

The applicability of segment 2 is determined by where the three straight lines described by the three asymptotes intersect in the $\log(W)$-$\log({R})$ plane. In particular, as justified in detail in Appendix~\ref{Sec:Appendix1}, one of the following conditions must be met to make segment 2 applicable: 
\begin{subequations}
\begin{align}
     0< B_{\mathrm z,2} < 1 ,\  R_{\mathrm z,1,2}(\theta) < R_{\mathrm z,1,3}(\theta),   \label{eq:condition_z1} \\
     B_{\mathrm z,2} > 1 ,\  R_{\mathrm z,1,2}(\theta) > R_{\mathrm z,1,3}(\theta),  \label{eq:condition_z2}
\end{align}
\end{subequations} 
where $ R_{\mathrm z,1,2}(\theta)$ and $R_{\mathrm z,1,3}(\theta)$ (expressions given in Table~\ref{table1}) 
are the critical distances at which segment 1 intersects segments 2 and 3, respectively. From the curves of $ R_{\mathrm z,1,2}(\theta)$ and $R_{\mathrm z,1,3}(\theta)$ plotted in Fig.~\ref{fig_parametersZ}~(b) we see that they intersect at a critical angle $\theta_{\mathrm z,2}$,  $R_{\mathrm z,1,3}(\theta)< R_{\mathrm z,1,2}(\theta)$ for $\theta \in (0, \theta_{\mathrm z,2})$, and $R_{\mathrm z,1,3}(\theta)> R_{\mathrm z,1,2}(\theta)$ for $\theta \in (\theta_{\mathrm z,2}, \frac{\pi}{2})$. The equation $R_{\mathrm z,1,2}(\theta) = R_{\mathrm z,1,3}(\theta)$ translates to 
\begin{align}
\label{eq:condition_z}
1 - \eta^2(\theta) = (2\sin^2\theta \cos\theta)^{B_{\mathrm z,2}(\theta)}, 
\end{align}
which, unfortunately, does not lead to a simple explicit expression for its root (i.e., for $\theta_{\mathrm z,2}$). Solving \eqref{eq:condition_z} numerically, we obtain $\theta_{\mathrm z,2} \approx 0.3285\pi > \theta_{\mathrm z,1}$.

Based on the above discussion, we conclude that for $\theta\in [\theta_{\mathrm z,1}, \theta_{\mathrm z,2}] \cup \{ \frac{\pi}{2} \}$, the asymptotic function $\Tilde{W}_{\mathrm z}(R;\theta)$ is dual-slope, formed by segments 1 and 3 with respective applicable $R$ ranges separated by critical distance $R_{\mathrm z,1,3}(\theta)$; whereas for $\theta \in (0, \theta_{\mathrm z,1}) \cup (\theta_{\mathrm z,2},\frac{\pi}{2})$, $\Tilde{W}_{\mathrm z}(R;\theta)$ is triple-slope, formed by all three segments with respective applicable $R$ ranges separated by $R_{\mathrm z,1,3}(\theta)$ and $R_{\mathrm z,2,3}(\theta)$, where $R_{\mathrm z,2,3}(\theta)$ is the critical distance at which segments 2 and 3 intersect. 
The expression of $R_{\mathrm z,2,3}(\theta)$ is also given in Table~\ref{table1}, and its curve plotted in Fig.~\ref{fig_parametersZ}~(b). Note that the expression of $R_{\mathrm z,2,3}(\theta)$ is not defined for $\theta =\theta_{\mathrm z,1}$ and $\theta = \frac{\pi}{2}$, for which segments 2 and 3 are parallel to each other. Finally, we recall the condition $R \gtrsim 10\lambda$ to ensure the validity of all asymptote segments.

\subsection{Asymptotic expressions for \texorpdfstring{$\Tilde{W}_{\mathrm x}(R;\theta)$}{x}}
\label{sec:3b}

\begin{figure}[!t]
\centering
\subfigure[$B_{\mathrm x,2}(\theta)$]{\includegraphics [width= .88\linewidth,trim= 30 280 30 290, clip]{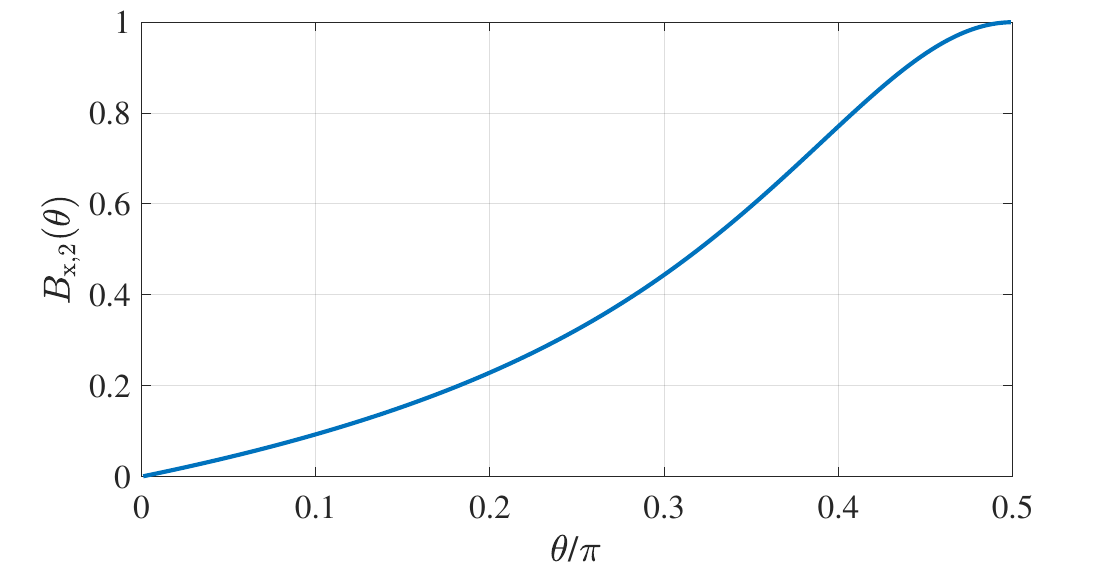} } 
\subfigure[Critical distances]{\includegraphics [width= .88\linewidth,trim= 30 280 30 290, clip]{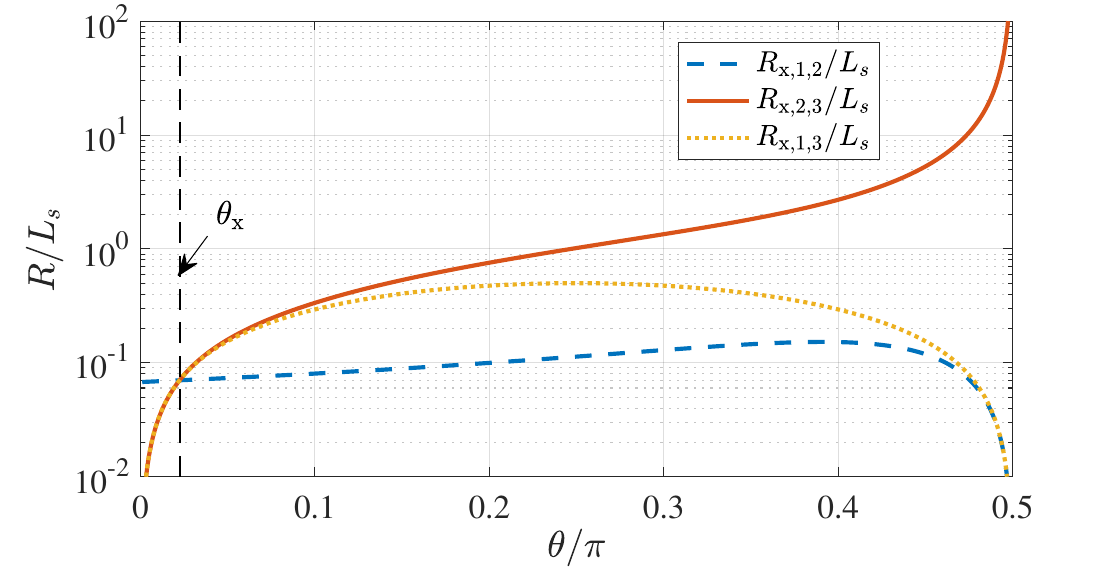} }\vspace{-.5em}
\caption{SBE $B_{\mathrm x,2}(\theta)$ and critical distances (divided by $L_s$) for $\Tilde{W}_{\mathrm x}(R;\theta)$; with $\theta_{\mathrm x} \approx 0.0225 \pi$.} 
\vspace{-1em}
\label{fig_parametersX}
\end{figure}

For $\theta \neq \frac{\pi}{2}$, the asymptotic analysis in Appendix \ref{Sec:Appendix2} results in three asymptotes for $\Tilde{W}_{\mathrm x}(R;\theta)$ of the form \eqref{eq:asymptote_general}, for small, medium, and large $R$ regimes respectively. They are summarized in Table~\ref{table2} and referred to as segments 1, 2, and 3. The SBEs of segments 1 and 3, $B_{\mathrm x,1}=0$ and $B_{\mathrm x,3}=1$, are the same as their counterparts for $\Tilde{W}_{\mathrm z}(R;\theta)$. The SBE of segment 2 is given by
\begin{align}\label{eq:beta}
    B_{\mathrm x,2}(\theta)  = \frac{1}{2} \left[  \eta^2(\theta)  + \eta(\theta) \right], 
\end{align}
where $\eta(\theta)$ is given by \eqref{eq:eta_1}. The curve of $B_{\mathrm x,2}(\theta)$ is plotted in Fig.~\ref{fig_parametersX}~(a) for $\theta \in (0, \frac{\pi}{2})$. As can be seen, $B_{\mathrm x,2}(\theta)$ increases gradually from $0$ to $1$ with $\theta$ in this range. Since $B_{\mathrm x,2}(\theta) <1$ always holds, the condition corresponding to \eqref{eq:condition_z1} should be met to make segment 2 applicable.

The critical radial distances at which each pair of segments intersect are given by $R_{\mathrm x,1,2}(\theta)$, $R_{\mathrm x,1,3}(\theta)$, and $R_{\mathrm x,2,3}(\theta)$ in Table~\ref{table2}, and their curves are plotted in Fig.~\ref{fig_parametersX}~(b) for $\theta \in (0, \frac{\pi}{2})$. 
We can see that the condition $R_{\mathrm x,1,2}(\theta) < R_{\mathrm x,1,3}(\theta)$ is violated when $\theta \leq \theta_{\mathrm x}$, where the critical angle $\theta_{\mathrm x}$ can be found by solving equation $R_{\mathrm x,1,2}(L,\theta) = R_{\mathrm x,1,3}(L,\theta)$, which translates to 
\begin{align}\label{eq:condition_x}
1 - \eta(\theta) = (2\sin\theta \cos^2\theta)^{B_{\mathrm x,2}(\theta)}.
\end{align}
Again, no simple expression has been found. Solving the equation numerically, we obtain $\theta_{\mathrm x} \approx 0.0225 \pi$. 
We can now conclude that for $\theta\in (0, \theta_{\mathrm x}]$, the asymptotic expression for $\Tilde{W}_{\mathrm x}(R;\theta)$ is dual-slope, formed by segments 1 and 3 with their respective applicable $R$ ranges separated by critical distance $R_{\mathrm x,1,3}(\theta)$; whereas for $\theta \in  (\theta_{\mathrm x},\frac{\pi}{2})$, the asymptotic expression is triple-slope, formed by all three segments with their respective applicable $R$ ranges separated by critical distances $R_{\mathrm x,1,3}(\theta)$ and $R_{\mathrm x,2,3}(\theta)$.

For the special case\footnote{This case is special because the second expression in \eqref{eq:wx0}, $ w_{\mathrm x2}(l; \mathbf{\Omega})$, will never apply no matter how large $R$ becomes.} of $\theta = \frac{\pi}{2}$, we obtain in Appendix~\ref{Sec:Appendix2} a different asymptote for the large $R$ regime, denoted by $\tilde{W}_{\mathrm x}^{(3^*)}(R;\theta)$, referred to as segment 3$^*$ in Table~\ref{table2}. No applicable asymptote for medium $R$ is obtained. This leads to a dual-slope asymptotic function, formed by segments 1 and 3$^*$ with respective applicable $R$ ranges separated by the critical distance $R_{\mathrm x,1,3^*}(\theta) = \frac{L_s}{\sqrt{8}}$.

We also remark that for those $\theta$ values close to $\frac{\pi}{2}$, the triple-slope asymptotic function formed by segments 1, 2, and 3 will lead to large errors in medium $R$ regime (see Appendix~\ref{Sec:Appendix2} for an explanation). Nevertheless, we do not aim for more accurate asymptotic expression for these $\theta$ conditions since they admit much smaller spatial bandwidth than those favorable angles (near $\frac{\pi}{4}$) for the same $R$. In other words, one should avoid the combination of $\theta \simeq \frac{\pi}{2}$ and $\hatbf{v}=\hatbf{e}_{\mathrm x}$ for good spatial DOF performance. Finally, we note the condition $R\geq R_{\mathrm x,0}$ (see \eqref{eq:R0}) to ensure the validity of all the derived asymptotes.

\subsection{Goodness-of-fit evaluation} 

\begin{figure}[!t]
\centering 
\subfigure[$\lambda \tilde{W}_{\mathrm z}(R;\theta)$ v.s.  $\lambda {W}_{\mathrm z}(R;\theta)$]{\includegraphics[height= 4.7cm,trim= 40 250 60 270, clip]{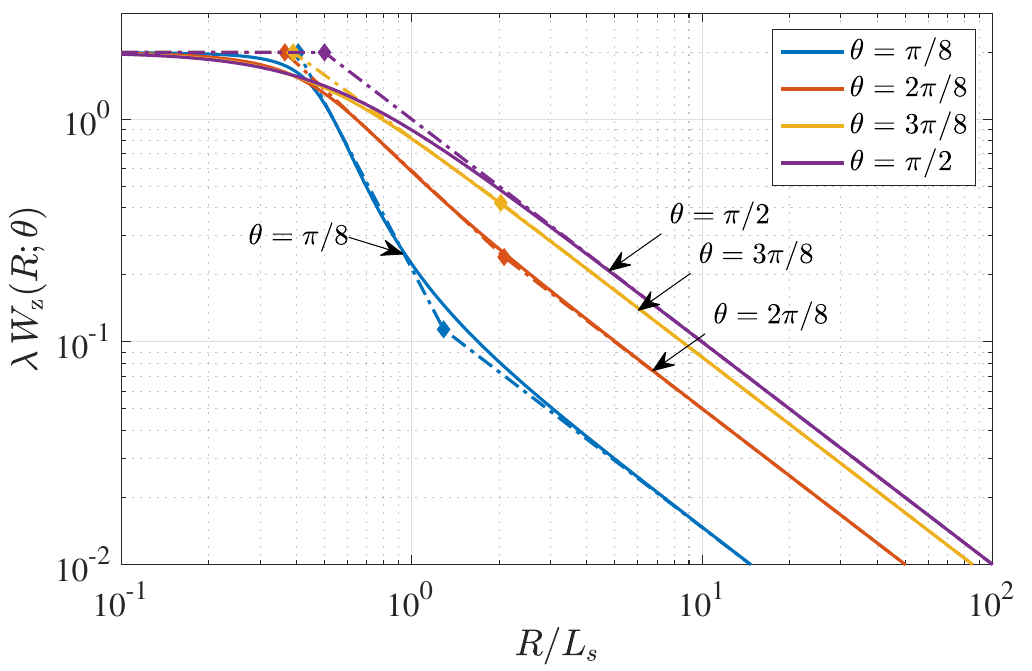}  }  
\subfigure[$\lambda \tilde{W}_{\mathrm x}(R;\theta)$  v.s. $\lambda {W}_{\mathrm x}(R;\theta)$]{\includegraphics [height= 4.7cm,trim= 40 250 60 270, clip, clip]{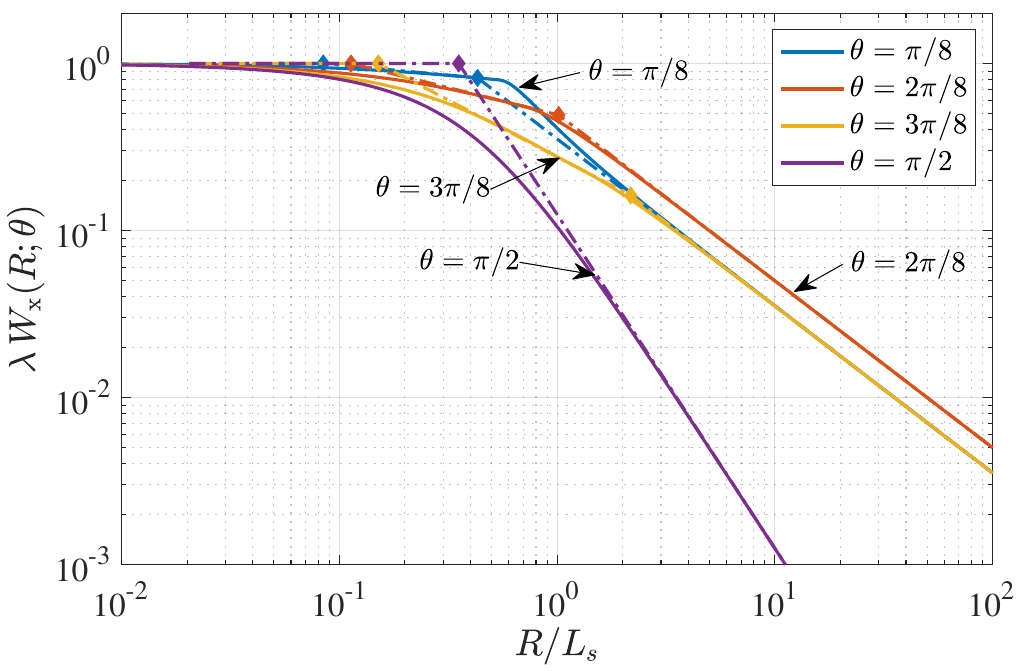} }
\caption{Log-domain comparison of the multi-slope asymptotic spatial bandwidth results (dash-dotted line) and exact results (solid line). Turning points of the asymptotic curves at critical distances are depicted using diamond markers.
$L_s = 1000 \lambda$ is assumed to ensure the applicability of the asymptotic expressions.}\vspace{-1em}
\label{fig:AsymptptesXZ}
\end{figure}

\begin{figure}[!t]
\centering 
\subfigure[$(\tilde{W}_{\mathrm z} - W_{\mathrm z})/{W_{\mathrm z}}$]{\includegraphics[height= 5.5cm,trim= 20 220 40 245, clip]{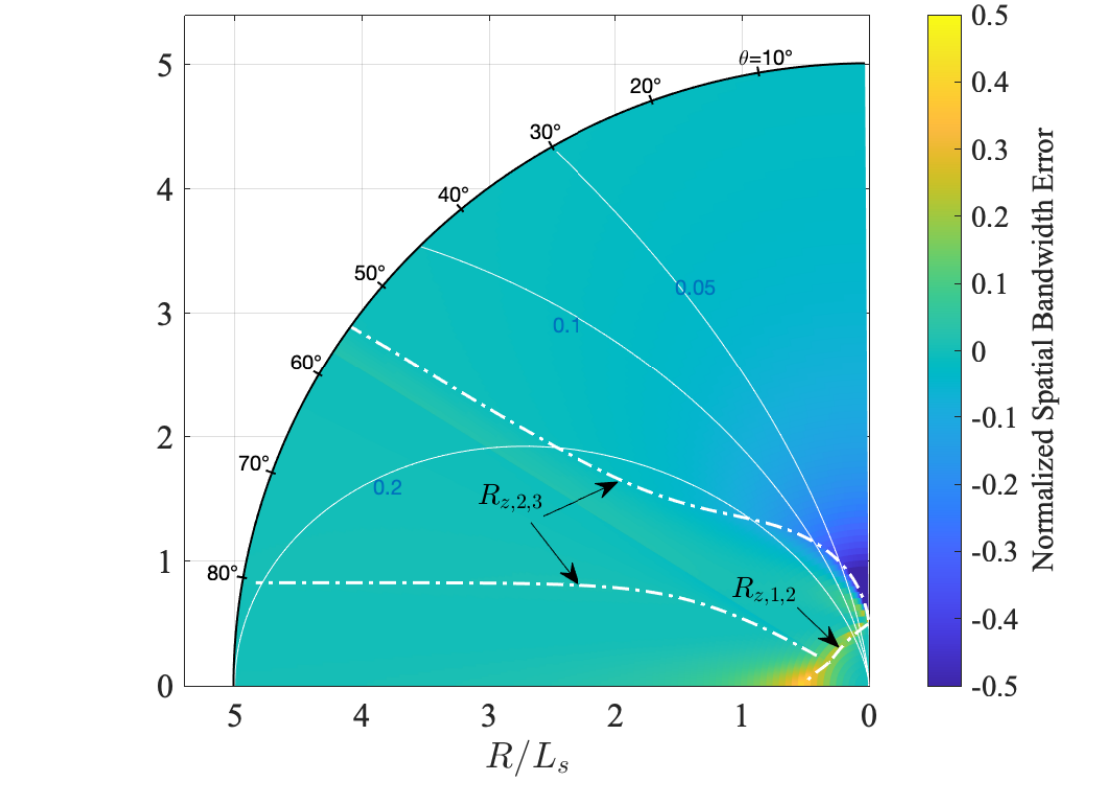}  } 
\subfigure[$(\tilde{W}_{\mathrm x} - W_{\mathrm x})/{W_{\mathrm x}}$]{ \includegraphics [height= 5.5cm,trim= 20 220 40 245, clip]{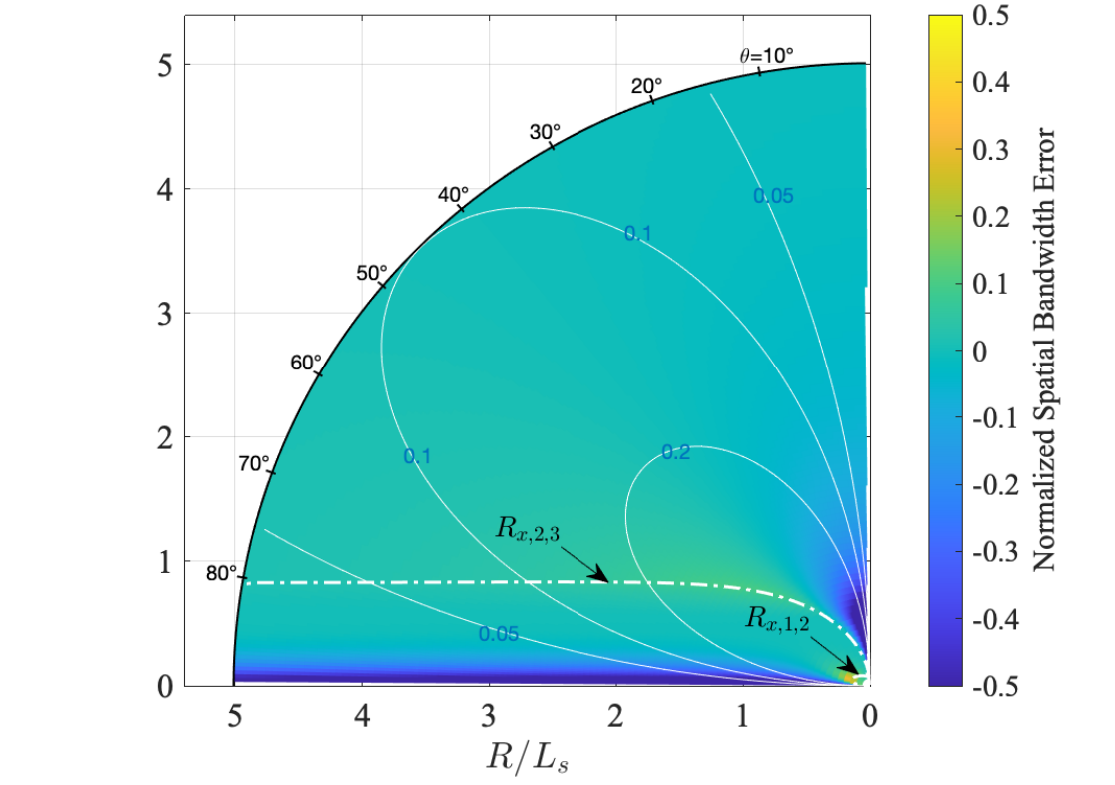}  }  
\caption{Normalized spatial bandwidth approximation errors caused by the derived multi-slope asymptotic functions. Contour lines of $\lambda W_{\mathrm z}$ and $\lambda W_{\mathrm x}$ at $0.2$, $0.1$, and $0.05$ (white solid thin lines, obtained approximately using the last asymptote segments) show that the $\theta$ ranges near $\frac{\pi}{2}$ and near $\frac{\pi}{4}$ admit larger spatial bandwidths for the two orientations respectively.}
\vspace{-1em}
\label{fig:Asymptpte_error}
\end{figure}

\begin{figure}[!t]
\centering 
\subfigure[$(\tilde{W}_{\mathrm z}^{\mathrm{ds}} - W_{\mathrm z})/{W_{\mathrm z}}$]{\includegraphics [height= 5.5cm,trim= 20 220 40 245, clip]{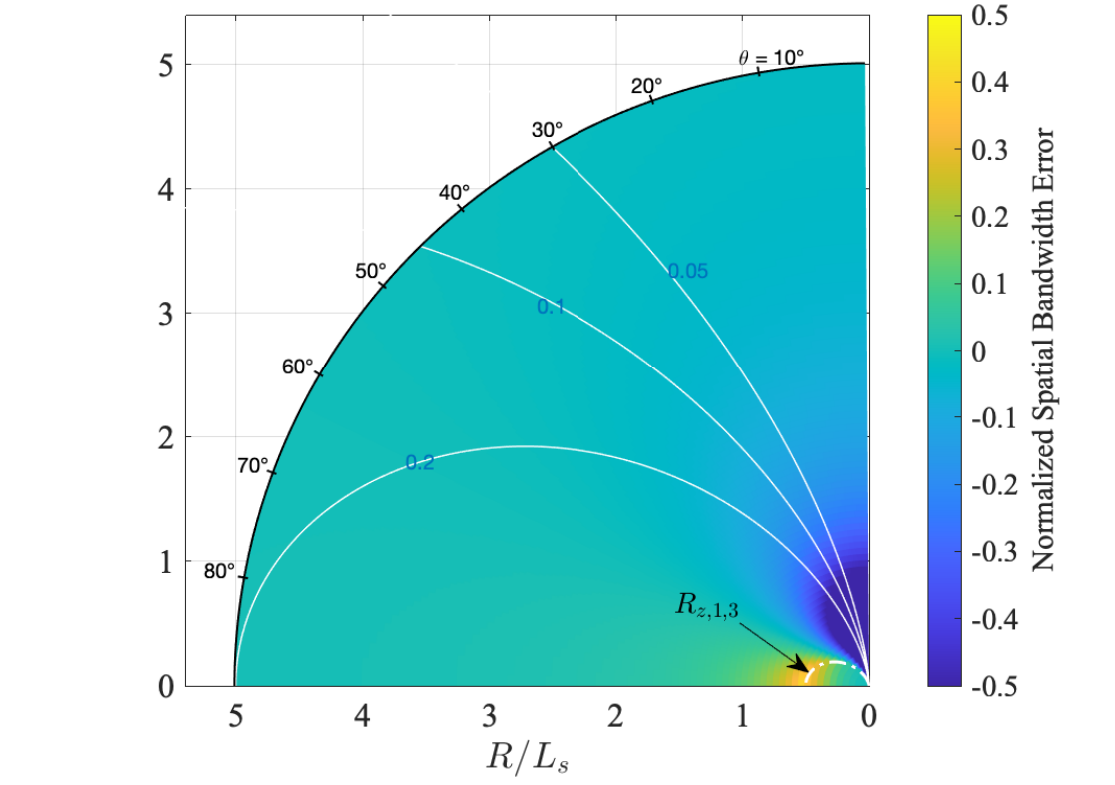} }
\subfigure[$(\tilde{W}_{\mathrm x}^{\mathrm{ds}} - W_{\mathrm x})/{W_{\mathrm x}}$]{ \includegraphics [height= 5.5cm,trim= 20 220 40 245, clip]{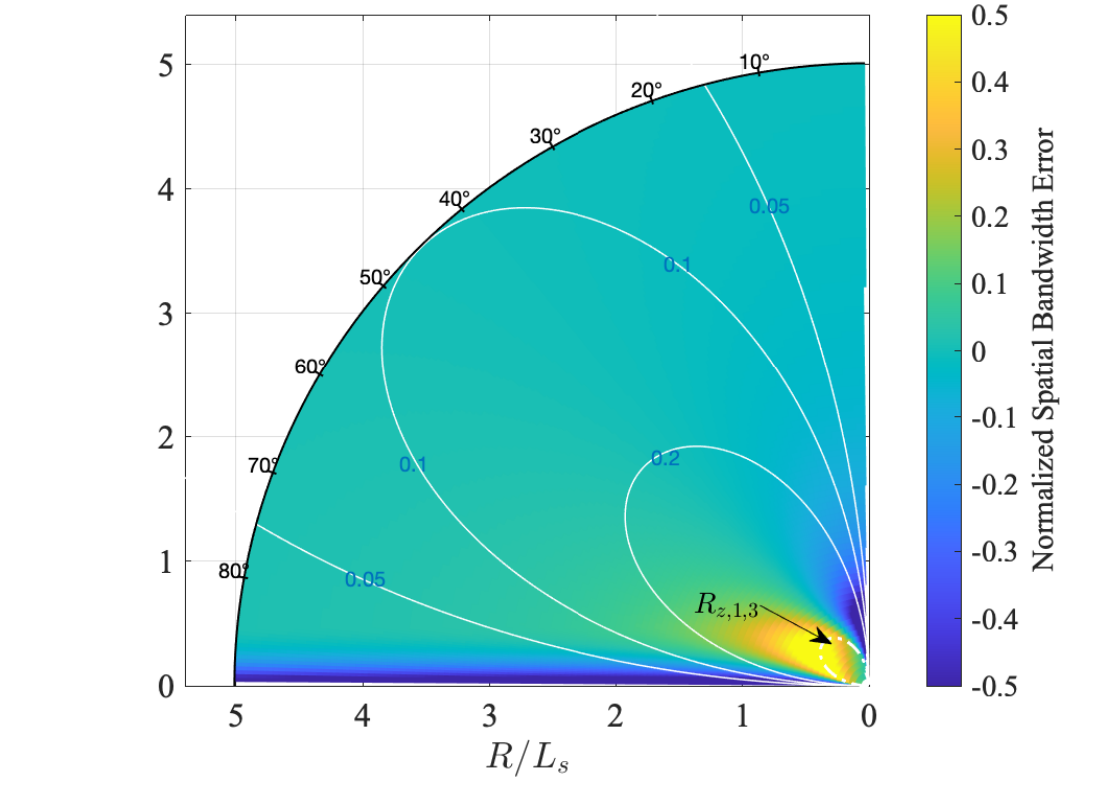}  }  
\caption{Normalized spatial bandwidth approximation errors caused by the dual-slope asymptotic functions (formed using segments 1 and 3 for both orientations, labeled using superscript `$\mathrm{ds}$') applied to the entire $\theta$ range.} 
\vspace{-0.5em}
\label{fig:Asymptpte_error_ds}
\end{figure}

In Fig.~\ref{fig:AsymptptesXZ}, we compare the log-log plots of $\lambda \, \tilde{W}_{\mathrm z}(R;\theta)$ and $\lambda \, \tilde{W}_{\mathrm x}(R;\theta)$ against ${R}/{L_s}$ (the results are frequency independent in this way) with their exact counterparts computed based on \eqref{eq:wz0} and \eqref{eq:wx0} for four evenly-spaced $\theta$ values. It can be seen that the asymptotic curves fit the exact results well for these $\theta$ values. In Fig.~\ref{fig:Asymptpte_error}, the colormaps of the normalized approximation errors, $(\tilde{W}_{\mathrm z} -W_{\mathrm z})/W_{\mathrm z }$ and $(\tilde{W}_{\mathrm x} - W_{\mathrm x})/W_{\mathrm x}$, are displayed in the $R$-$\theta$ plane for $\theta\in(0,\frac{\pi}{2}]$ and $R\leq 5L_s$. As noted already,  the asymptotic expression $\tilde W_{\mathrm x}(R;\theta)$ fit badly when $\theta$ is near $\frac{\pi}{2}$. Otherwise, larger errors occur only for small $R$, either near the critical distances (plotted in white dash-dotted lines) or when $\theta$ is small. For those $\theta$ values that admit a larger spatial bandwidth (near $\frac{\pi}{2}$ for $W_{\mathrm z}$ or near $\frac{\pi}{4}$ for $W_{\mathrm x}$), the asymptotic expressions exhibit a good fit. 

From Fig.~\ref{fig:AsymptptesXZ} and Fig.~\ref{fig:Asymptpte_error}, we observe that the last asymptote segments, which have SBE $1$, fit the exact curve tightly as soon as $R$ exceeds a few $L_s$. We also note that, although the critical distances $R_{\mathrm z,2,3}$ and $R_{\mathrm x,2,3}$ can be very large under certain $\theta$ conditions, the SBEs of the second asymptotes $B_{\mathrm z,2}(\theta)$ and $B_{\mathrm x,2}(\theta)$ are nevertheless close to $1$ under the same conditions, as can be verified from Fig.~\ref{fig_parametersZ} and Fig.~\ref{fig_parametersX}. Fig.~\ref{fig:Asymptpte_error_ds} presents the colormaps of the normalized computation error caused by the dual-slope expressions (formed by segments 1 and 3 (or 3$^*$)) applied to the entire $(0,\frac{\pi}{2}]$ range. Compared to Fig.~\ref{fig:Asymptpte_error}, the additional large errors only occur for small $R$.

\section{Dual-Slope Asymptotic Function for the General Orientation}
\label{sec:4}

The above observations motivate us to obtain a dual-slope asymptotic expression for $W_{\hatbf v}(R;\theta) \triangleq w_{\hatbf{v}}(0,\mathbf{\Omega})$, for an arbitrary orientation $\hatbf v = (\hat{v}_{\mathrm x}, \hat{v}_{\mathrm y}, \hat{v}_{\mathrm z} )^{\mathrm T}  \neq \hatbf{e}_{\mathrm y}$. This is done through a geometric interpretation of its asymptotic behavior for very small and very large $R$. 

\begin{figure}[!t]
\centering 
	\includegraphics [width= \linewidth,trim= 0 0 0 0, clip]{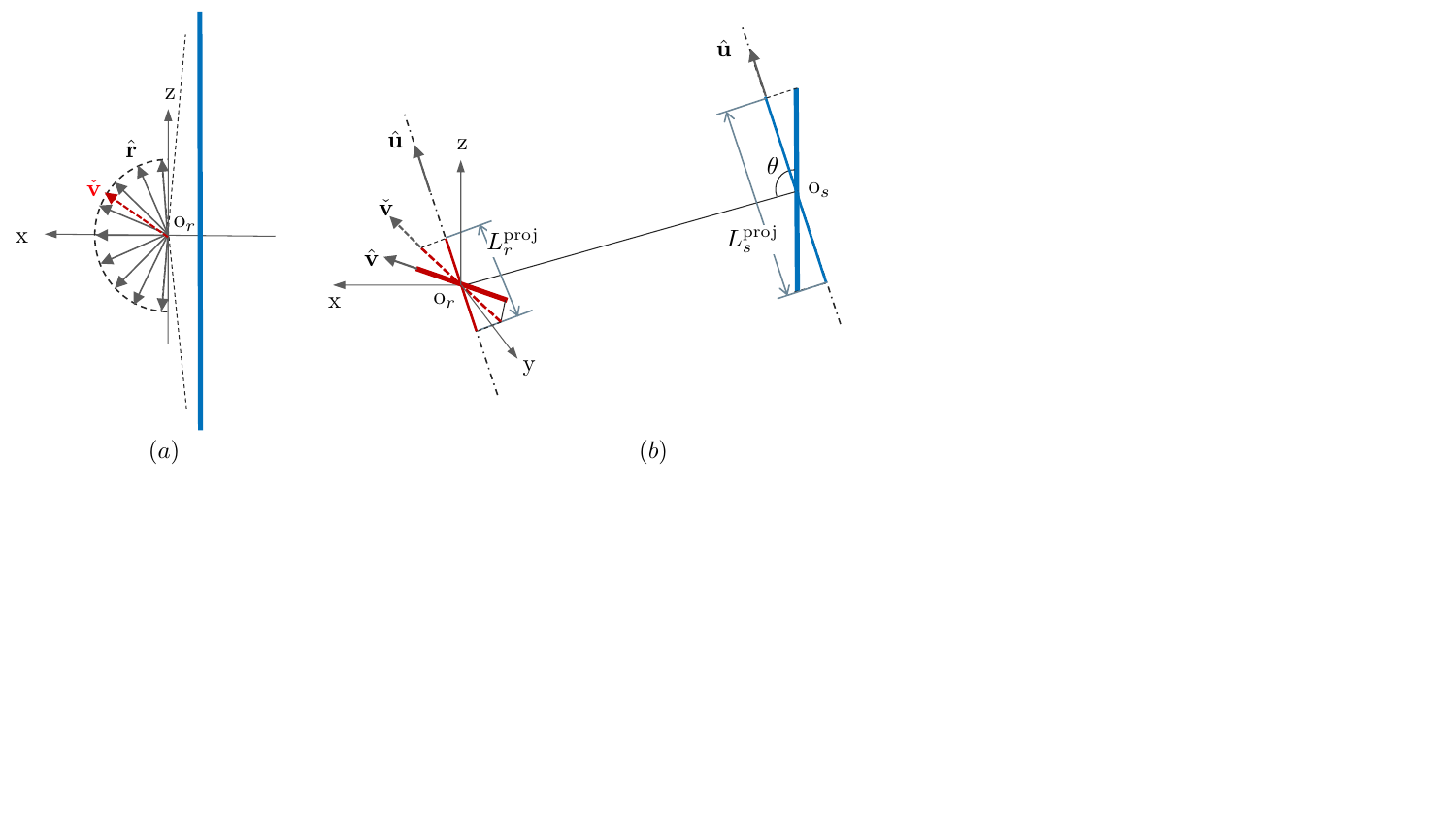} \vspace{-1em}
	\caption{The geometric interpretation of the dual-slope asymptotic analysis for arbitrary orientation $\hatbf v$. (a) As $R\rightarrow 0^+$, the set of $\hatbf r$ asymptotically forms a semi-ring of unit radius on the $+\mathrm{x}$-$\mathrm{z}$ half-plane; (b) For large $R$, it is equivalent to consider the pair of perfectly aligned linear arrays given by projecting $\mathcal L_s$ and $\mathcal L_r$ in the direction of $\hatbf{u} \triangleq (-\cos\theta,  0, \sin\theta)^{\mathrm T}$. 
 }  \vspace{-1.5em}
	\label{fig:asymp_geo_any}
\end{figure}

When $R$ is small, we expect $W_{\hatbf v}(R;\theta)$ to be asymptotically constant. As $R \rightarrow 0^+$, $\mathcal L_r$ is very close to the center of $\mathcal L_s$ for any $\theta$.  Due to the large $L_s$ assumption, asymptotically, the set $\{\hatbf r(0,q)\}_{|q|\leq \frac{L_s}{2}}$ forms a semiring of unit radius on the $+\mathrm{x}$-$\mathrm{z}$ half-plane, as shown in Fig.~\ref{fig:asymp_geo_any}~(a). As a result, the maximum spatial frequency $\kappa_{\hatbf v}^{\mathrm{max}} = \frac{1}{\lambda}  \| \check{\mathbf v}\|  =\frac{1}{\lambda} \, \sqrt{ \hat{v}_{\mathrm x}^2 + \hat{v}_{\mathrm z}^2}$ is achieved by $\hatbf r = \frac{\check{\mathbf v}} {\| \check{\mathbf v} \|}$, where $\check{\mathbf v} \triangleq (\hat{v}_{\mathrm x}, 0, \hat{v}_{\mathrm z})^\mathrm{T}$ is the projection of $\hatbf v$ on the $\mathrm{x}$-$\mathrm{z}$ plane; whereas the minimum spatial frequency $\kappa_{\hatbf v}^{\mathrm{min}} = - |\hat{v}_{\mathrm z}| $ is caused when $\hatbf r = -\mathrm{sign}(\hat{v}_{\mathrm z} ) \,\hatbf e_{\mathrm z}$. Following the definition spatial bandwidth \eqref{eq:spatial_bandwidth_RCS}, we obtain
\begin{equation}\label{eq:14}
    W_{\hatbf v}^c (R) \sim
    \kappa_{\hatbf v}^{\mathrm{max}} - \kappa_{\hatbf v}^{\mathrm{min}} = \frac{1}{\lambda} \, \big(\sqrt{ \hat{v}_{\mathrm x}^2 + \hat{v}_{\mathrm z}^2} + |\hat{v}_{\mathrm z}| \big) \quad (R \rightarrow 0^+). 
\end{equation}

When $R$ is very large, the expressions of $\tilde{W}_{\mathrm z}^{(3)}(R;\theta)$ and $\tilde{W}_{\mathrm x}^{(3)}(R;\theta)$ given in Table~\ref{table1} and Table~\ref{table2} lead to the same DOF results as the paraxial approximation does. For $\hatbf v = \hatbf{e}_{\mathrm z}$, we have that $\widetilde{K}_{\mathrm z}  =  \frac{L_s \sin\theta}{\lambda R} \cdot L_r \sin\theta$. We see that $L_s \sin\theta$ and $L_r \sin\theta$ are the projected lengths of  $\mathcal L_s$ and $\mathcal L_r$ onto the line on the $\mathrm{x}$-$\mathrm{z}$-plane that is perpendicular to the $\mathrm o_s$-$\mathrm o_r$ connecting line. The direction of this line is given by the unit vector $\hatbf u \triangleq (-\cos\theta,  0, \sin\theta)^{\mathrm T}$, as shown in Fig.~\ref{fig:asymp_geo_any}~(b). Also, for  $\hatbf v = \hatbf{e}_{\mathrm x}$,  we have that $\widetilde{K}_{\mathrm x}  =  \frac{L_s \sin\theta}{\lambda R} \cdot L_r |\cos\theta|$, where  $L_s \sin\theta$ and  $L_r |\cos\theta|$ are the projected lengths of $\mathcal L_s$ and $\mathcal L_r$ onto $\hatbf u$.  Hence, $\widetilde{K}_{\mathrm z}$ and $\widetilde{K}_{\mathrm x}$ are the results given by the paraxial approximation, as discussed in the Section~\ref{sec:1}. We use this equivalence to obtain the asymptote for $W_{\hatbf v}(R;\theta)$ for large $R$. However, it is important to note that in the specific case of $\hatbf v = \hatbf{e}_{\mathrm x}$ and $\theta = \frac{\pi}{2}$, the projection method leads to $\widetilde{K}_{\mathrm x} = 0$, which clearly is incorrect. Our analysis has identified this discrepancy and provides a separate expression $\tilde{W}_{\mathrm x}^{(3*)}(R;\theta)$ for this case, which exhibits a decay rate twice as large as the others (i.e., SBE 2). The correct SBE cannot be found by just changing the direction of the projection.

The projected length of $\mathcal L_s$ depends only on $\theta$, and hence, we denote by $L_{s}^{\mathrm{proj}}(\theta) = L_s \sin\theta$. The projected length of $\mathcal L_r$ depends additionally on its orientation, and we denote it by $L_{r}^{\mathrm{proj}}(\theta,\hatbf{v})$. Using $\hatbf v = (\hat{v}_{\mathrm x}, \hat{v}_{\mathrm y}, \hat{v}_{\mathrm z} )^{\mathrm T} $, its explicit expression can be derived: 
\begin{equation}
    L_r^{\mathrm{proj}}(\theta,\hatbf{v}) =L_r |\langle \hatbf v, \hatbf u \rangle | = L_r  |\hat{v}_{\mathrm x} \cos\theta   -\hat{v}_{\mathrm z} \sin\theta |. 
\end{equation}
Replacing $L_s$ and $L_r$ in $\frac{L_s L_r}{\lambda R}$ using $L_{s}^{\mathrm{proj}}$ and $L_{r}^{\mathrm{proj}} (\theta,\hatbf{v})$ respectively and divide the results by $L_r$, we obtain the following asymptote: 
\begin{equation}\label{eq:17}
    W_{\hatbf v} (R; \theta) \sim \frac{1}{\lambda}\, |\hat{v}_{\mathrm x} \cos\theta   -\hat{v}_{\mathrm z} \sin\theta  | \sin\theta \cdot  \frac{L_s}{R} \quad (R\to \infty).
\end{equation}

\begin{table*}[!t]
\caption{Dual-slope asymptotic function $\tilde{W}_{\hatbf{v}}(R;\theta)$, valid when $\hatbf v \neq \hatbf e_{\mathrm y}$ and $R >  \left( \frac{L_r}{2} +10\lambda \right)  \frac{1}{\sin\theta}$}
\label{table3}
\centering \vspace{-.5em}
\small
\renewcommand{\arraystretch}{1.4}
\begin{tabular}{l l}
\begin{tabular}{|p{.52\linewidth}|}
\hline 
\textbf{Expression } \\
\hline
$\Tilde{W}_{\hatbf v}(R;\theta) = \begin{cases}
		\frac{1}{\lambda}\, \big( \sqrt{ \hat{v}_{\mathrm x}^2 + \hat{v}_{\mathrm z}^2} + |\hat{v}_{\mathrm z}| \big)  ,  & R \in (0,  R_{\hatbf v} ] , \\
		\frac{1}{\lambda}\,  |\hat{v}_{\mathrm x} \cos\theta   -\hat{v}_{\mathrm z} \sin\theta | \sin\theta \cdot \frac{L_s}{R} ,  & R \in (R_{\hatbf v}, \infty ).
	\end{cases}$\vspace{1pt} \\
\hline
\end{tabular}
& \hspace{-1.4em}
\begin{tabular}{|p{.4\linewidth}|} 
\hline
\textbf{Critical distances} \\ \hline
\vspace{1.1pt} 
$R_{\hatbf v} =  \frac{L_s |\hat{v}_{\mathrm x} \cos\theta   -\hat{v}_{\mathrm z} \sin\theta  | \sin\theta }{\sqrt{ \hat{v}_{\mathrm x}^2 + \hat{v}_{\mathrm z}^2} + |\hat{v}_{\mathrm z} | }$ \\[4pt]
\hline
\end{tabular}
    \end{tabular}\vspace{-1em}
\end{table*}

The critical distance $R_{\hatbf v}$, given in Table~\ref{table3}, at which the two asymptotes \eqref{eq:14} and \eqref{eq:17} intersect, can be easily computed. This completes all the information needed to form the dual-slope asymptotic expression for $\Tilde{W}_{\hatbf v}(R;\theta)$ for $\theta\in (0,\pi)$, as summarized in Table~\ref{table3}. By substituting $\hatbf v= \hatbf e_{\mathrm z}$ or $\hatbf v= \hatbf e_{\mathrm x}$, it can be easily verified that the resultant dual-slope asymptotic expressions are exactly the same as the two formed using asymptotes 1 and 3 for $\mathrm z$ and $\mathrm x$ directions.

Finally, we  note that $R_{\hatbf v} \leq \frac{L_s}{2}$ holds for any given $\theta$ and $\hatbf v$. For $\hat{v}_{\mathrm x} =0$, the proof is trivial and hence omitted. For $\hat{v}_{\mathrm x} \neq 0$, we let $\zeta = \arctan({\hat{v}_{\mathrm z}}/{\hat{v}_{\mathrm x}})$, which makes $\hat v_{\mathrm z} = \| \check{\mathbf v}\| \sin\zeta$ and $\hat v_{\mathrm x} = \| \check{\mathbf v}\| \cos\zeta$. Since $\| \check{\mathbf v}\|  =\sqrt{ \hat{v}_{\mathrm x}^2 + \hat{v}_{\mathrm z}^2} > 0$ (since $\hatbf{v}\neq \hatbf{e}_{\mathrm y}$) and $\sin\theta > 0$, $R_{\hatbf v}$ can be rewritten as 
\begin{align*}
    R_{\hatbf v} &= \frac{L_s |\cos\zeta \cos\theta   -\sin\zeta \sin\theta| \sin\theta }{1 + |\sin\zeta| } \nonumber \\
    &= \frac{L_s |\cos(\zeta +\theta)| \sin\theta }{1 + |\sin \zeta|} 
    = \frac{L_s | \sin(\zeta + 2\theta) - \sin\zeta| }{ 2( 1 + |\sin\zeta|)} \leq \frac{L_s}{2}, 
\end{align*}
where equality is achieved when $ \sin(\zeta + 2 \theta) =1$ if $\sin\zeta \leq 0$, or when $ \sin(\zeta + 2 \theta) =-1$ if $\sin \zeta > 0$.

\section{Spatial Multiplexing Region Evaluation}
\label{sec:5}

In this section, we demonstrate the usage of the asymptotic expression given in Table~\ref{table3} in the evaluation of \textit{spatial multiplexing region} in a simple communication scenario under random orientation conditions of the receiving array.
 
\subsection{Performance metrics}
\label{sec:5a}

The \textit{spatial multiplexing region} \cite[Definition 3]{Ding2022degrees}, as a fundamental measure of the DOF performance of the \ac{LOS} channel between two antenna arrays, is defined as the set of locations surrounding $\mathcal L_s$ where the K number given by \eqref{eq:spatial_DoF_RCS} exceeds a given threshold $K_0$, indicating that the required spatial multiplexing capability can be met. The definition is given with respect to a specific orientation of the receiving array. In what follows, we first extend the definition to arbitrary orientation conditions. To make the definitions concrete, we let $(\varphi, \gamma)$ be the pair of zenith and azimuth angles that parameterize $\hatbf{v}$ and denote their valid value range by $\mathcal O$, which is a subset of $[0, \pi]\times [0,2\pi)$. We note the following: first, the reference coordinate system for $(\varphi, \gamma)$ can be arbitrary; secondly, $(\varphi, \gamma)$ can be treated as random variables, and thirdly, $\mathcal O$ can have a dependency on the location of $\mathcal L_r$.  

\begin{figure}[!t]
	\centering
	\includegraphics [width= .64\linewidth]{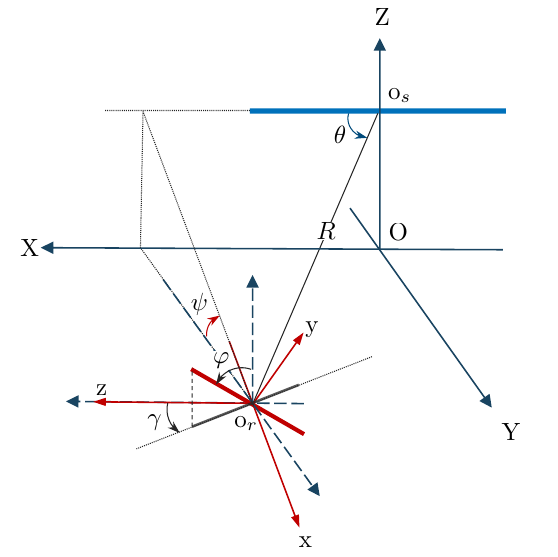} \vspace{-1em}
	\caption{Scenario illustration for the case study.}  \vspace{-1em}
	\label{fig:Knumber_scenario}
\end{figure}

\begin{definition}[Maximum Spatial Multiplexing Region]
\label{Def:Max_sm_region}

Given a pair of linear arrays $(\mathcal L_s, \mathcal L_r)$ and the valid value range $\mathcal{O}$ for $(\varphi, \gamma)$,  the maximum spatial multiplexing region, denoted by $\mathcal R^{\max}( K_0)$, is the set of positions in the 3D space, where when $\mathrm o_r$ is placed, the maximum value of K number achievable in the \ac{LOS} channel between $\mathcal L_s$ and $\mathcal L_r$ is equal to or greater than a given threshold $K_0$. Namely, 
	\begin{equation}\label{eq:max_sm_region}
		\mathcal R^{\max}( K_0)	\triangleq 
		\left\{ (R,\theta): \max\nolimits_{(\varphi, \gamma) \in \mathcal{O}} \left\{ K_{\hatbf v}(R,\theta) \right\} \geq K_0 \right\}. 
	\end{equation} 
\end{definition}

\begin{definition}[Expected Spatial Multiplexing Region]
\label{Def:sm_region1}
Given a pair of linear arrays $(\mathcal L_s, \mathcal L_r)$ and a joint probability distribution of $(\varphi, \gamma)$ over the valid value range $\mathcal{O}$, denoted by $p_{\varphi,\gamma}(\varphi,\gamma; \mathcal O)$, the expected spatial multiplexing region, denoted by ${\mathcal R}^{\mathrm{exp}}(K_0)$, is the set of positions in the 3D space, where when $\mathrm o_r$ is placed, the expected value of the achievable K number in the \ac{LOS} channel between $\mathcal L_s$ and $\mathcal L_r$ is equal to or greater than a given threshold $K_0$. Namely, 
	\begin{equation}\label{eq:sm_region}
		{\mathcal R}^{\mathrm{exp}}(K_0)	\triangleq 
		\left\{ (R,\theta): \mathbb{E}_{p_{\varphi,\gamma}(\varphi,\gamma; \mathcal O)} \left\{ K_{\hatbf v}(R,\theta) \right\} \geq K_0 \right\}.
	\end{equation} 
\end{definition}

We remark that each pair of $(R,\theta)$ values describes a circle surrounding $\mathcal L_s$, and therefore, $\mathcal R_{\mathcal O}^{\max}( K_0)$ and $\Bar{\mathcal R}_{\mathcal O}(K_0)$ are 3D regions that are rotationally symmetric around $\mathcal L_s$. They are the sets of locations where a satisfactory DOF can be achieved in the expectation and optimal sense. 

\subsection{Scenario description}
\label{sec:5b}

We consider a simple scenario shown in Fig.~\ref{fig:Knumber_scenario}, and adopt a global coordinate system (GCS) $\mathrm X$-$\mathrm Y$-$\mathrm Z$, whose $\mathrm X$-$\mathrm Y$ plane is regarded the ground plane, for its description. The source array $\mathcal L_s$ is placed parallel to the $\mathrm X$-axis at a height $Z_s > \frac{L_s}{2}$. Its center is then given by $\mathrm o_s = (0,0,Z_s)^{\mathrm T}$. The location of the receiving array $\mathcal L_r$ is restricted in the $\mathrm X$-$\mathrm Y$ plane. In particular, $\mathrm o_r = (X_r,Y_r,0)^{\mathrm T}$, where $Y_r \geq 0$, is assumed. It is easy to obtain 
\begin{align}\label{eq:R_and_theta}
 R = \sqrt{Z_s^2 + X_r^2 + Y_r^2}, \quad   \theta = \arccos\left(\frac{X_r}{R} \right).
\end{align}

We use the zenith and azimuth angles $(\varphi,\gamma)$ defined in the GCS, as shown in Fig.~\ref{fig:Knumber_scenario}, to specify the orientation of $\mathcal L_r$. The unit directional vector in the GCS is hence given by  
\begin{align}\label{eq:vhat_GCS}
   \hatbf v_{\mathrm{G}} =  (\sin\varphi \cos\gamma ,  \sin\varphi \sin\gamma , \cos\varphi)^{\mathrm T}.
\end{align} 
Denote the angle between the $\mathrm X$-$\mathrm Y$ plane and the $\mathcal L_s$-$\mathrm o_r$ plane by $\psi$. The above notations readily lead to $\cos\psi = \frac{Y_r}{\sqrt{Z_s^2+ Y_r^2}}$ and $\sin \psi = \frac{ Z_s}{\sqrt{Z_s^2+ Y_r^2}}$. Following the definition of the \ac{LCS} given in Section~\ref{sec:2a}, it is not difficult to verify that the transformation between GCS and LCS can be performed using a rotation matrix $ \mathbf Q = [ 0 , \cos\psi , -\sin \psi ; 0 , \sin \psi , \cos\psi; 1, 0, 0]$. Specifically, the unit directional vector is given by $\hatbf v = \mathbf Q \hatbf v_{\mathrm{G}}$ in the LCS.

We consider two orientation conditions for $\mathcal{L}_r$, representing two different potential application scenarios. In the first, $\mathcal{L}_r$ can rotate freely in 3D, so $(\varphi, \gamma) \in\mathcal O_\mathrm{3D} = [0, \pi]\times [0,2\pi)$, which can happen on a handheld terminal. In the second, the rotation of $\mathcal{L}_r$ is restricted in the $\mathrm X$-$\mathrm Y$ plane, i.e., $(\varphi, \gamma) \in\mathcal O_\mathrm{2D} = \{ \frac{\pi}{2} \}\times [0,2\pi)$, which may happen if mounted on a car roof. To compute the expected spatial multiplexing region, we assume that the orientation is uniformly distributed. Specifically, the \textit{3D uniform orientation}, denoted by $\mathrm{uni3D}$, assumes that $\hatbf v$ is uniformly distributed over the unit sphere centered at $\mathrm o_r$. That is, $(\varphi,\gamma)$ follows the joint \ac{PDF} $p_{\varphi,\gamma}^{\mathrm{uni3D}}(\varphi,\gamma) = p_{\gamma}(\gamma) p_{\varphi}(\varphi)$, where 
\begin{subequations}
     \begin{align}
        p_{\gamma}(\gamma) &= 
        \frac{1}{2\pi}, \quad \text{for } 0\leq \gamma < 2\pi,  \label{eq:22}\\
        p_{\varphi}(\varphi) &= 
        \frac{\sin\varphi}{2}, \quad \text{for }  0\leq \varphi \leq \pi.  \label{eq:23}
    \end{align}   
\end{subequations}
The \textit{2D uniform orientation}, denoted by $\mathrm{uni2D}$,  assumes that $\hatbf v$ is uniformly distributed over the unit circle centered at $\mathrm o_r$ in the $\mathrm X$-$\mathrm Y$ plane. That is, $(\varphi,\gamma)$ follows the joint \ac{PDF} $p_{\varphi,\gamma}^{\mathrm{uni2D}}(\varphi,\gamma) = p_{\gamma}(\gamma) \delta(\varphi - \frac{\pi}{2})$ where $p_{\gamma}(\gamma)$ is given in \eqref{eq:22} and $\delta(\cdot)$ is the Dirac delta function.

\subsection{Expected and maximum K number}
\label{sec:5c}

We now derive the maximum and expected values of the K number following \eqref{eq:7} and Table~\ref{table3}. Recall that the critical distance $R_{\hatbf v}$ given in Table~\ref{table3} is upper bounded by $\frac{L_s}{2}$. Since $Z_s > \frac{L_s}{2}$ is assumed, we adopt the approximation 
\begin{equation}\label{eq:K22}
	\widetilde{K}_{\hatbf v}(R, \theta) =  
		 |\hat{v}_{\mathrm x} \cos\theta  -\hat{v}_{\mathrm z} \sin\theta | \sin\theta \, \frac{L_s L_r}{\lambda R}
\end{equation}
for any location of $\mathcal L_r$. 
Given $\mathrm o_r$, $\theta$ is determined, and the analysis is performed only on $|\hat{v}_{\mathrm x} \cos\theta   -\hat{v}_{\mathrm z} \sin\theta |$. 

\subsubsection{Maximum values}

Recall that $|\hat{v}_{\mathrm x} \cos\theta   -\hat{v}_{\mathrm z} \sin\theta | \equiv |\langle \hatbf v, \hatbf u \rangle |$, where $\hatbf u = (-\cos\theta, 0, \sin\theta)^{\mathrm T}$. Given the 3D orientation freedom, \eqref{eq:K22} is maximized when $\hatbf v= -\mathrm{sign}(\cos\theta ) \,\hatbf u$, such that $|\hat{v}_{\mathrm x} \cos\theta   -\hat{v}_{\mathrm z} \sin\theta | =1$. Consequently, an approximation of $ \max\nolimits_{(\varphi, \gamma) \in \mathcal{O}_{\mathrm{3D}}} \left\{ K_{\hatbf v}(R,\theta) \right\}$ is  
\begin{equation}\label{eq:K_opt_3D}
 \widetilde{K}^{\max}_{\mathrm{3D}}(R,\theta)
    = \frac{\sin\theta L_s L_r}{\lambda R}.
\end{equation}

When $\mathcal L_r$ is restricted to be in the $\mathrm X$-$\mathrm Y$ plane, the maximum value of $|\langle \hatbf v, \hatbf u \rangle |$ is achieved when $\hatbf v$ aligns with the projection of $\hatbf u$ in the $\mathrm X$-$\mathrm Y$ plane. Since $\hatbf u_{\mathrm{G}} = \mathbf Q^\mathrm{T} \hatbf u  = (\sin\theta, -\cos\theta \cos\psi, \cos\theta \sin\psi)^{\mathrm T}$, it is computed that  
\begin{align*}
\max\nolimits_{(\varphi, \gamma) \in \mathcal{O}_{\mathrm{2D}}} 
\big\{ |\langle \hatbf v, \hatbf u \rangle | \big\} 
&= \sqrt{ \sin^2\theta + \cos^2\theta \cos^2\psi } \nonumber\\
&= \sqrt{1 - \cos^2\theta \sin^2\psi} 
\end{align*}
Accordingly, an approximation of $ \max\nolimits_{(\varphi, \gamma) \in \mathcal{O}^{\mathrm{2D}}} \left\{ K_{\hatbf v}(R,\theta) \right\}$ 
is given by 
\begin{align}\label{eq:K_opt_2D}
    \widetilde K_{\mathrm{2D}}^{\max}(R,\theta) 
    &= \sqrt{1 - \cos^2\theta \sin^2\psi} \, \frac{\sin\theta L_s L_r}{\lambda R}.
\end{align}

From \eqref{eq:K_opt_3D} and \eqref{eq:K_opt_2D} we can see that $\widetilde K_{\mathrm{uni2D}}^{\max}(R,\theta) \leq \widetilde K_{\mathrm{uni3D}}^{\max}(R,\theta)$ always holds. This is expected since $\mathcal O_\mathrm{3D}$ offers more freedom for orientation control. The gap between $\widetilde K_{\mathrm{uni2D}}^{\max}(R,\theta)$ and $\widetilde K_{\mathrm{uni3D}}^{\max}(R,\theta)$ shrinks as $\psi \rightarrow 0$ (e.g., when $Y_r$ is large), which can also be reasoned from geometry: the $\mathrm x$-$\mathrm z$ plane of LCS, which dominates the contribution in spatial DOF, approaches the $\mathrm X$-$\mathrm Y$ plane of GCS as $\psi \rightarrow 0$. 

\begin{remark}
The above discussion suggests two low-complexity near-optimal orientation control strategies for $\mathcal L_r$ based on its position under the 3D and 2D orientation constraints, respectively. However, from the discussion in Section \ref{sec:3}, we know that when $\mathcal L_r$ and $\mathcal L_s$ are very close ($Z_s$ needs to be small), it is best to always align them in parallel regardless of the actual value of $\theta$. Following the same approach as we identify the critical distances for the asymptotic functions, we suggest $ R_{\mathrm{3D}}(\theta) =  \frac{L  \sin\theta }{2}$ and $R_{\mathrm{2D}}(\theta) =  \sqrt{1 - \cos^2\theta \sin^2\psi}\frac{\sin\theta  L}{2}$ as the distance thresholds that divide the small and large $R$ regimes for these orientation control strategies to apply. 
\end{remark}

\subsubsection{Expected values} 

Under the assumption of 3D uniform orientation, the statistics of $|\hat{v}_{\mathrm x} \cos\theta -\hat{v} \sin\theta |$ is independent of $\psi$, which can be verified based on the rotational symmetry of the distribution of $\hatbf v$  around the $\mathrm{z}$-axis at $\mathrm{o}_r$. Therefore, we can simplify the analysis by letting $\psi = 0$, which leads to $\hatbf v =  \mathbf Q \hatbf v_{\mathrm{G}} =  (\sin\varphi \sin\gamma, \cos\varphi, \sin\varphi \cos\gamma)^{\mathrm T}$. The expected value of $|\hat{v}_{\mathrm x} \cos\theta -\hat{v}_{\mathrm z} \sin\theta |$ is thus given by
\begin{align*}
&\mathbb E_{p_{\varphi,\gamma}^{\mathrm{uni3D}}} \left\{ |\hat{v}_{\mathrm x} \cos\theta   -\hat{v} \sin\theta | \right\} \nonumber \\
 = &  \frac{1}{4\pi}  \int_0^{2\pi} \hspace{-3pt} \int_0^{\pi} \!| \sin\varphi \sin\gamma \cos\theta  - \sin\varphi \cos\gamma \sin\theta|\, \sin\varphi \, \mathrm d \varphi  \mathrm d \gamma \\
 = & \frac{1}{4\pi}  \int_0^{2\pi} | \sin\gamma \cos\theta  - \cos\gamma \sin\theta| \int_0^{\pi} \sin^2\varphi \, \mathrm d \varphi \mathrm d \gamma   =\frac{1}{2}. 
\end{align*}
Accordingly, an approximation of $\mathbb{E}_{p_{\varphi,\gamma}^{\mathrm{uni3D}}} \left\{ K_{\hatbf v}(R,\theta) \right\}$ is obtained: 
\begin{equation}\label{eq:K_expect_uni3D}
\widetilde{K}^{\mathrm{exp}}_{{\mathrm{uni3D}}}(R, \theta) 
= \frac{1}{2} \frac{\sin\theta L_s L_r}{\lambda R}. 
\end{equation}

Under the assumption of 2D uniform orientation, $\hatbf v = \mathbf Q \hatbf v_{\mathrm{G}} = ( \cos\psi \sin\gamma, \sin \psi \sin\gamma, \cos\gamma )^{\mathrm T}$ is obtained by substituting $\varphi = \frac{\pi}{2}$ into \eqref{eq:vhat_GCS}. The expected value of $|\hat{v}_{\mathrm x} \cos\theta -\hat{v} \sin\theta |$ can be computed
\begin{align*}
&\mathbb E_{p_{\varphi,\gamma}^{\mathrm{uni2D}}}\left\{ |\hat{v}_{\mathrm x} \cos\theta   -\hat{v}_{\mathrm z} \sin\theta | \right\}  \nonumber \\
= &\frac{1}{2\pi}  \int_0^{2\pi} |\cos\psi \sin\gamma \cos\theta \!- \! \cos\gamma \sin\theta| \,  \mathrm d \gamma  \nonumber \\
= &\frac{2}{\pi} \sqrt{1 - \cos^2\theta \sin^2\psi}.
\end{align*}  
Consequently, an approximation of $\mathbb{E}_{p_{\varphi,\gamma}^{\mathrm{uni2D}}} \left\{ K_{\hatbf v}(R,\theta) \right\}$ is given by  
\begin{equation}
\label{eq:K_expect_uni2D}
\widetilde{K}^{\mathrm{exp}}_{{\mathrm{uni2D}}}(R, \theta) 
= \frac{2}{\pi} \sqrt{1 - \cos^2\theta \sin^2\psi} \, \frac{\sin\theta L_s L_r }{\lambda R}. 
\end{equation}

From \eqref{eq:K_expect_uni3D} and \eqref{eq:K_expect_uni2D}, we notice that $\widetilde{K}^{\mathrm{exp}}_{{\mathrm{uni2D}}}(R, \theta)$ can be greater or smaller than $\widetilde{K}^{\mathrm{exp}}_{{\mathrm{uni3D}}}(R, \theta)$ depending on $\theta$ and $\psi$. When $\psi \rightarrow 0$, $\sqrt{1 - \cos^2\theta \sin^2\psi} \rightarrow 1$, and $\widetilde{K}^{\mathrm{exp}}_{{\mathrm{uni2D}}}(R, \theta)  \rightarrow \frac{4}{\pi} \widetilde{K}^{\mathrm{exp}}_{{\mathrm{uni3D}}}(R, \theta)$, showing that the 3D uniform orientation distribution leads to a K number performance penalty in the expectation sense.

\subsection{Spatial multiplexing regions on the ground plane}
\label{sec:5d}

We examine the intersections of spatial multiplexing regions with the $\mathrm X$-$\mathrm Y$ plane based on \eqref{eq:K_opt_3D}, \eqref{eq:K_opt_2D},  \eqref{eq:K_expect_uni3D}, and \eqref{eq:K_expect_uni2D}, since $\mathrm o_r$ is restricted in it. 
We begin with the expected spatial multiplexing region under the 3D uniform orientation assumption. 
Substituting $\sin\theta =  \sqrt{Z_s^2 + Y_r^2}/R$ and $\cos\theta = X_r/R$, which follows directly from \eqref{eq:R_and_theta}, into $\widetilde{K}^{\mathrm{exp}}_{{\mathrm{uni3D}}}(R, \theta)$ given by \eqref{eq:K_expect_uni3D}, the intersecting region can be written explicitly as follows: 
\begin{equation}\label{eq:SM_expected_3D}
	{\mathcal R}^{\mathrm{exp}}_{\mathrm{uni3D}}(K_0;Z_s)	= 
	\Big\{ (X_r,Y_r):  \frac{\sqrt{Z_s^2 + Y_r^2}}{2 R^2}  \geq \frac{1}{G_0} \Big\}, 
\end{equation} 
where $R = \sqrt{Z_s^2 + X_r^2 + Y_r^2}$, and $G_0 \triangleq \frac{L_s L_r}{\lambda K_0}$. The solution to the corresponding equation in the condition gives the boundary of this region, denoted by ${\mathcal B}^{\mathrm{exp}}_{\mathrm{uni3D}}(K_0;Z_s)$. After some simple algebra, it is found that ${\mathcal B}^{\mathrm{exp}}_{\mathrm{uni3D}}(K_0;Z_s)$ is given by the set of $(X_r,Y_r)$ satisfying 
\begin{align}\label{SM_expected_3Dboundary}
     X_r^2 =  \frac{G_0}{2} \sqrt{Z_s^2 + Y_r^2} - (Z_s^2 + Y_r^2), \quad 
     0\leq Y_r^2 \leq  \frac{G_0^2}{4} - Z_s^2. 
\end{align}
Thus, to ensure a nonempty ${\mathcal R}^{\mathrm{exp}}_{\mathrm{uni3D}}(K_0;Z_s)$, the condition $Z_s < G_0/2$ should be met.

From \eqref{eq:K_opt_3D} and \eqref{eq:K_expect_uni3D},  $\widetilde{K}_{\mathrm{3D}}^{\max}(R,\theta) = 2 \widetilde{K}_{\mathrm{uni3D}}^{\mathrm{exp}}(R, \theta)$ is observed. Therefore, replacing $K_0$ by $K_0/2$, \eqref{eq:SM_expected_3D} and \eqref{SM_expected_3Dboundary}  describe the intersection and its boundary of the maximum spatial multiplexing region with the $\mathrm X$-$\mathrm Y$ plane for the same threshold $K_0$. We denote them by 
\begin{align*}
{\mathcal R}^{\max}_{\mathrm{3D}}( K_0;Z_s) &= {\mathcal R}^{\mathrm{exp}}_{\mathrm{uni3D}}(K_0/2;Z_s), \;\text{and}
\nonumber \\
\mathcal B_{\mathrm{3D}}^{\max}(K_0;Z_s) &= {\mathcal B}^{\mathrm{exp}}_{\mathrm{uni3D}}(K_0/2;Z_s). 
\end{align*}

Next, we consider the expected spatial multiplexing region under the 2D uniform orientation assumption. By substituting $\sin\theta = \sqrt{Z_s^2 + Y_r^2}/R$ and $\cos\theta = X_r/R$ into \eqref{eq:K_expect_uni2D}, the intersection of $\widetilde{K}^{\mathrm{exp}}_{{\mathrm{uni2D}}}(R, \theta)$ with the $\mathrm X$-$\mathrm Y$ plane can be explicitly formulated  as follows: 
\begin{align}\label{eq:SM_expected_2D}
{\mathcal R}^{\mathrm{exp}}_{\mathrm{uni2D}}(K_0;Z_s)&= \nonumber \\ 
\Big\{(X_r,Y_r):& 
  \frac{2}{\pi} \frac{1}{R^2}  \sqrt{Z_s^2 + Y_r^2- \frac{ Z_s^2 X_r^2}{R^2} }   \geq \frac{1}{G_0}  \Big\}. 
\end{align} 
Its boundary, denoted by ${\mathcal B}_{\mathrm{uni2D}}^{\mathrm{exp}}(K_0; Z_s)$, can be found by solving the corresponding equation in the condition. The explicit expression for $(X_r, Y_r)$ forming the boundary is complicated in this case. Nevertheless, it can be easily verified that to ensure a nonempty ${\mathcal R}^{\mathrm{exp}}_{\mathrm{uni2D}}(K_0;Z_s)$, the condition $Z_s < \frac{2}{\pi}G_0$ should be met. 
From \eqref{eq:K_opt_2D} and \eqref{eq:K_expect_uni2D}, it can be seen that  $\widetilde{K}_{\mathrm{2D}}^{\max}(R,\theta) = \frac{\pi}{2} \widetilde{K}_{\mathrm{uni2D}}^{\mathrm{exp}}(R, \theta)$. Consequently, the intersection for the maximum spatial multiplexing region and the boundary are given by 
\begin{align*}
\mathcal R_{\mathrm{2D}}^{\max}( K_0; Z_s) &= \Bar{\mathcal R}_{\mathrm{uni2D}}(2K_0/\pi;Z_s), \;\text{and}
\nonumber \\ 
\mathcal B_{\mathrm{2D}}^{\max}( K_0; Z_s) &= \Bar{\mathcal B}_{\mathrm{uni2D}}(2K_0/\pi; Z_s). 
\end{align*}

\begin{figure}[!t]
	\centering
	\subfigure[3D orientation]{ \includegraphics [width= .9\linewidth,trim= 20 190 20 190, clip]{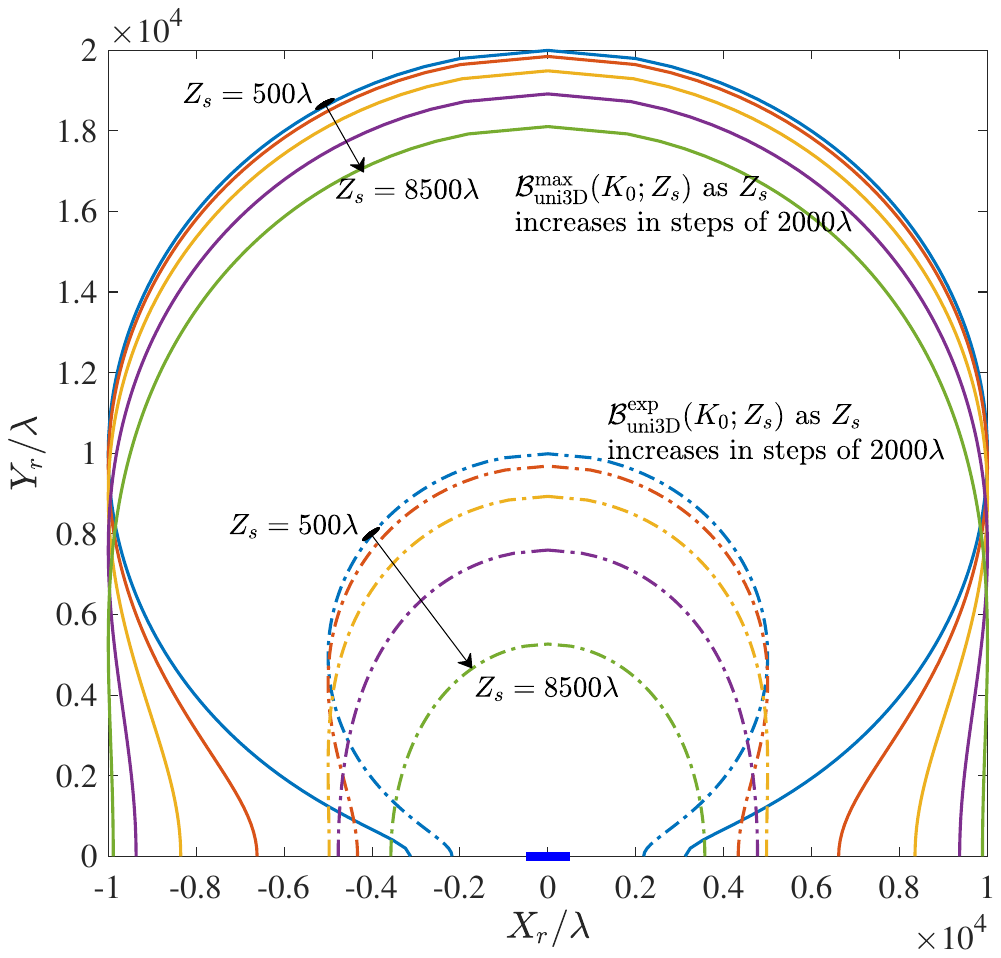 }} 
	\hfill
	\subfigure[2D  orientation]{\includegraphics [width= .9\linewidth,trim= 20 190 20 190, clip]{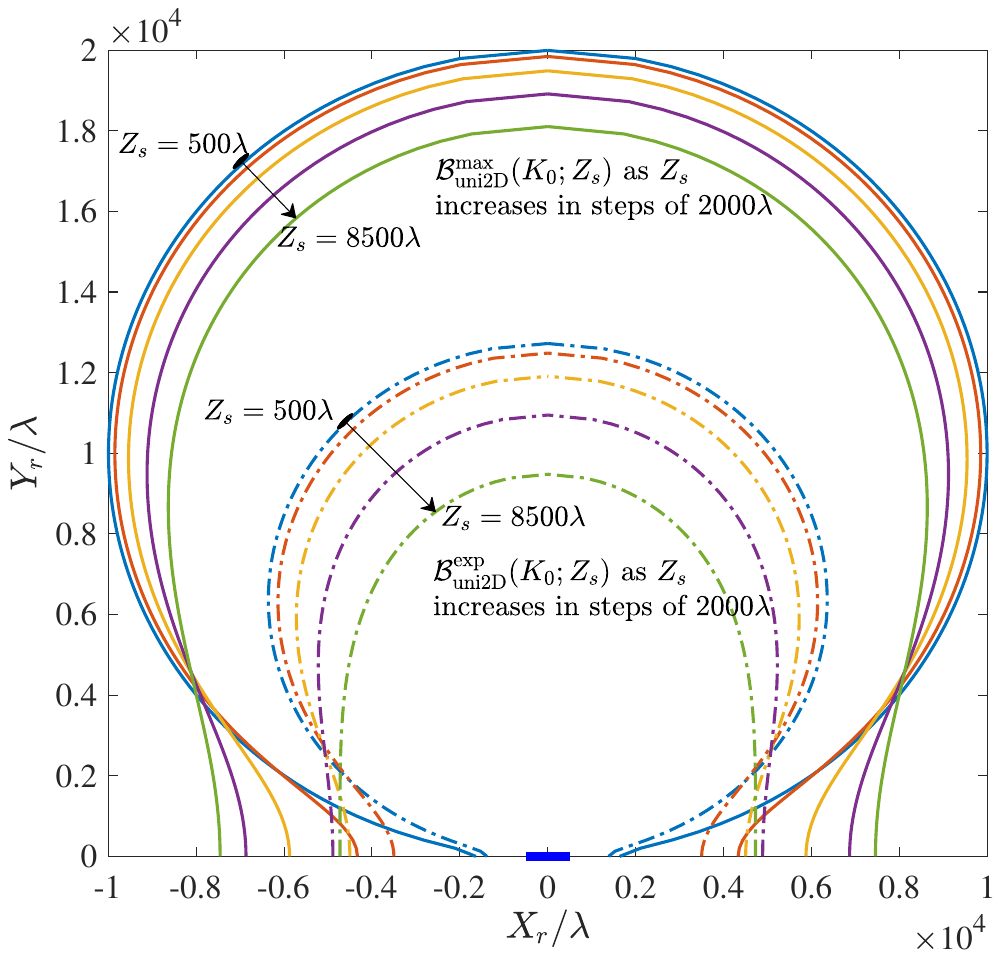} }
	\caption{Boundaries of the expected spatial multiplexing regions (dash-dotted lines) and the maximum spatial multiplexing regions (solid lines) on the $\mathrm{X}$-$\mathrm{Y}$-plane under the 3D and 2D orientation restrictions, with $L_s = 1000 \lambda$,  $L_r = 20 \lambda$, and $K_0 = 1$.  The short blue bar centered at $(0,0)$ depicts the source array, which is lifted by height $Z_s$ off the plane. }\vspace{-1em} 
 \label{fig:sm_region}
\end{figure}

In Fig.~\ref{fig:sm_region}, these boundaries are presented for the setting with $L_s = 1000\lambda$, $L_r=20\lambda$, and $K_0 = 1$. Five different values of $Z_s$, increasing from $500\lambda$ to $8500\lambda$ in steps of $2000\lambda$, are selected. We see that for any given $Z_s$, $\mathcal R_{\mathrm{3D}}^{\max}( K_0;Z_s)$ is always larger than $\mathcal R_{\mathrm{uni2D}}^{\max}( K_0;Z_s)$. This is expected since the 3D condition offers more freedom for orientation control than the 2D restriction. In contrast, $\mathcal R_{\mathrm{3D}}^{\mathrm{exp}}( K_0;Z_s)$ is much smaller than $\mathcal R_{\mathrm{uni2D}}^{\mathrm{exp}}( K_0;Z_s)$ for any given $Z_s$. This is also not a surprise (although it might seem so). Recall that the $\mathrm x$-$\mathrm z$ plane of the LCS dominates the contribution in spatial DOF, and this plane approaches the ground plane for large $Y_r$. Therefore, roughly speaking, restricting $\mathcal L_r$ to the ground plane actually helps to avoid many orientations that lead to a very small K number. Moreover, with optimal control, the difference between 3D and 2D orientation restrictions is more significant for locations closer to the $\mathrm{X}$ axis (with small $Y_r$); while in the expectation sense, the difference made by the 3D and 2D uniform orientation assumptions is more significant for locations further away from the $\mathrm{X}$ axis. Finally, we emphasize the impact of $Z_s$ on spatial multiplexing regions. Both the shape and the area of the coverage are affected.

\begin{figure}[!t]
 \centering 
\subfigure[$\widetilde{K}^{\max}_{\mathrm{3D}}(R,\theta)$ and $\widetilde K_{\mathrm{2D}}^{\max}(R,\theta)$]{\includegraphics [width= .9\linewidth,trim= 20 220 20 240, clip]{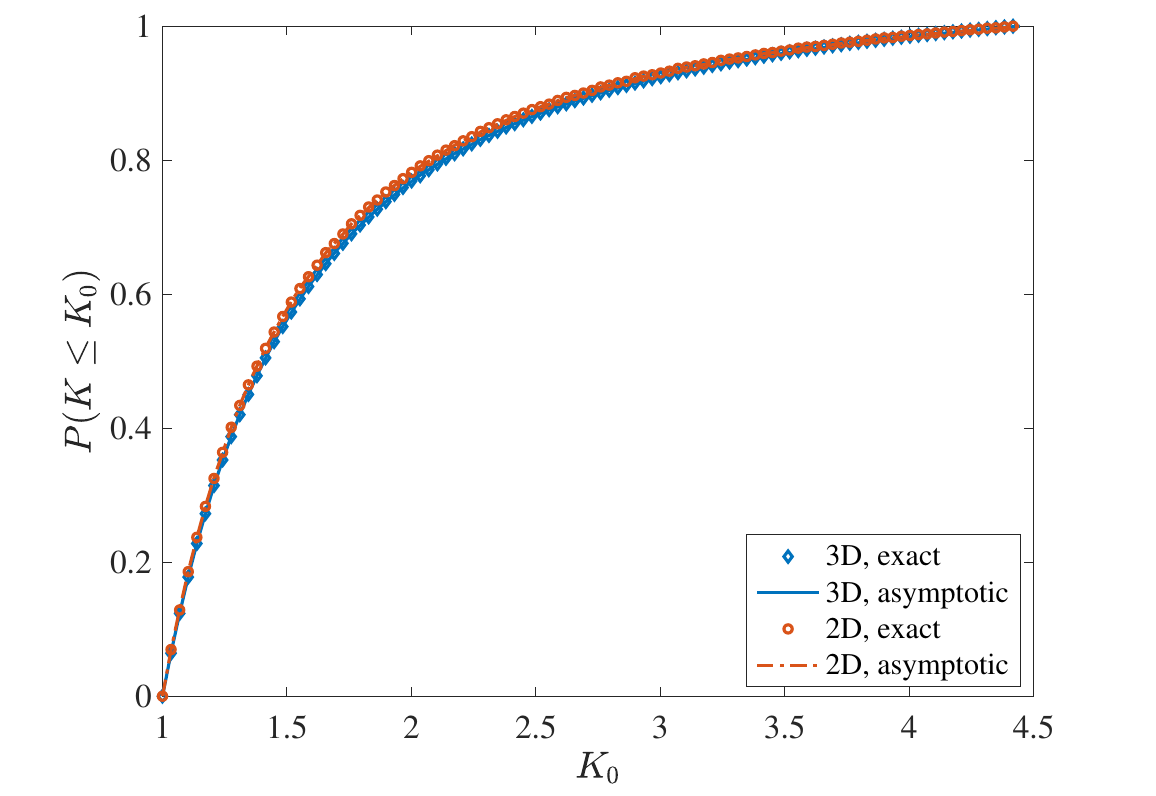} } 
\hfill
\subfigure[$\widetilde{K}^{\mathrm{exp}}_{{\mathrm{uni3D}}}(R, \theta) $ and 
$ \widetilde{K}^{\mathrm{exp}}_{{\mathrm{uni2D}}}(R, \theta)$]{ \includegraphics [width= .9\linewidth,trim= 20 220 20 240, clip]{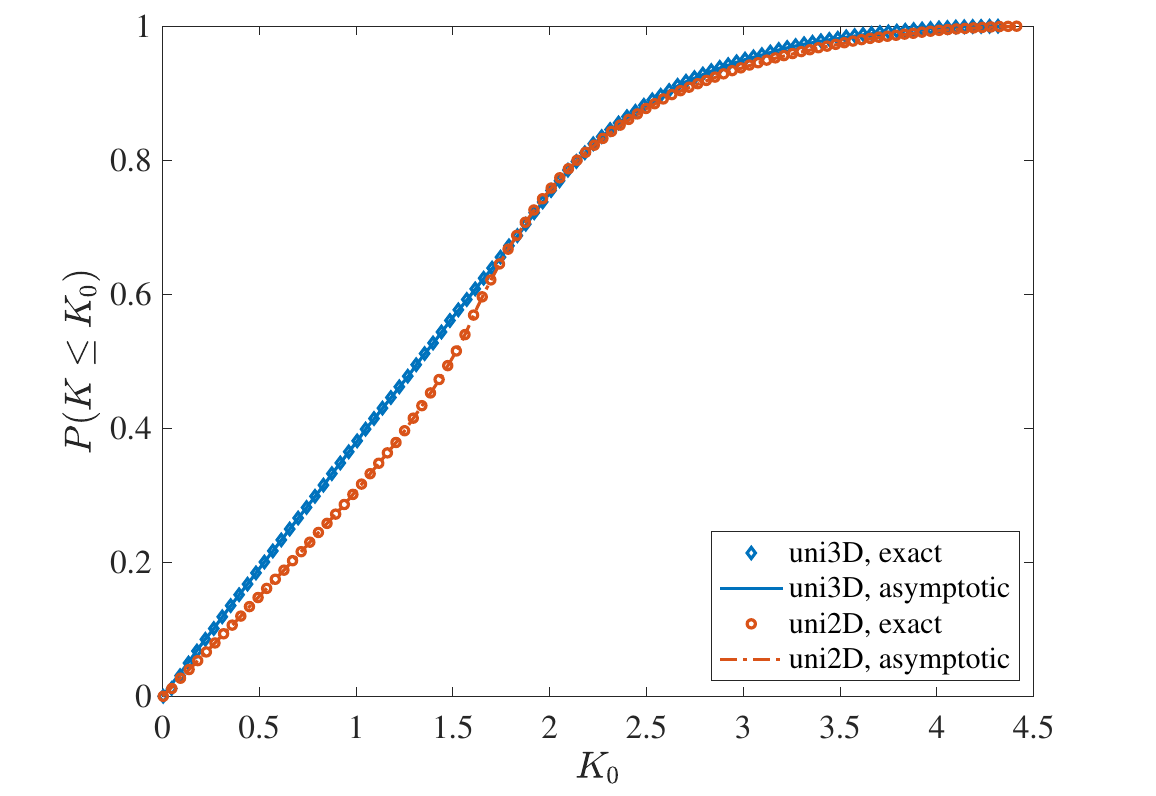}  } 
\caption{Simulated CDFs of the maximum and expected values of the K number when $\mathrm o_r$ is located inside the \textit{respective} spatial multiplexing regions. $L_s = 1000 \lambda$,  $L_r = 20 \lambda$,  $Z_s = 4500 \lambda$.} 
\vspace{-1em}
\label{fig:KCDF}
\end{figure}

The distributions of the maximum and expected values of the K number as the mobile user traverses the respective spatial multiplexing regions over grids of size $50\lambda \times 50 \lambda$ are also evaluated. In particular, the simulated \ac{CDF} curves of $\widetilde{K}^{\max}_{\mathrm{3D}}(R,\theta)$, $\widetilde K_{\mathrm{2D}}^{\max}(R,\theta)$, $\widetilde{K}^{\mathrm{exp}}_{{\mathrm{uni3D}}}(R, \theta) $, and 
$ \widetilde{K}^{\mathrm{exp}}_{{\mathrm{uni2D}}}(R, \theta)$ computed following \eqref{eq:K_opt_3D}, \eqref{eq:K_opt_2D}, \eqref{eq:K_expect_uni3D}, and \eqref{eq:K_expect_uni2D}, are compared with the results given by their exact counterparts computed using numerical integration based on \eqref{eq:spatial_bandwidth_RCS} and \eqref{eq:spatial_DoF_RCS}. 
$Z_s = 4500\lambda$ is chosen. 
For random orientation, $(\phi,\gamma)$ are generated following $p_{\varphi,\gamma}^{\mathrm{uni3D}}(\varphi,\gamma)$ and $p_{\varphi,\gamma}^{\mathrm{uni2D}}(\varphi,\gamma)$ described in Section~\ref{sec:5b}; the true optimal orientation is found by exhaustive search. The obtained \acp{CDF} are shown in Fig.~\ref{fig:KCDF}. A good match between the asymptotic and exact results can be seen in all cases, proving the goodness-of-fit of the asymptotic expression for the general $\hatbf v$ for this application. 
Combining Fig.~\ref{fig:KCDF} and Fig.~\ref{fig:sm_region}, we can conclude that with optimal orientation control, the 3D freedom results in a slightly better K number distribution in a larger spatial multiplexing region; while under the 3D and 2D uniform orientation conditions, the latter leads to a better distribution of the expected values of the K number in a significantly larger spatial multiplexing region.

\section{Conclusions}

Based on the exact closed-form expressions for the spatial bandwidth at the center of the receiving array, located in the radiative region of an \ac{LSAA} and orientated in $\hatbf{e}_{\mathrm x}$ and $\hatbf{e}_{\mathrm z}$ directions, we have derived asymptotic expressions of the form $W(R;\theta) =  A(\theta) \cdot ({L_s}/{R})^{B(\theta)}$, where $A(\theta)$ and $B(\theta)$ are piecewise constant in $R$, creating two or three asymptote segments depending on $\theta$. 
See Table~\ref{table1} and \ref{table2} for details.
When an expression has three asymptote segments, the middle segment ends at a distance that is at most a few times $L_s$. For this reason, we have also provided a two-segment asymptotic expression for the spatial bandwidth for an arbitrary orientation $\hatbf{v}$, at the expense of reduced accuracy for distances comparable to $L_s$, see Table~\ref{table3} for details. These expressions provide additional insight and detail beyond previously known results. For instance, the effect of orientation on the spatial bandwidth for very short distances is captured, the $\theta$- and orientation-dependent SBE of the middle segment is revealed, and the breakpoints of the distance ranges are made mathematically precise.

If $L_r$ is small relative to $R$, the product of $L_r$ and the spatial bandwidth provides an approximation of the available spatial DOF in the LOS channel. Under this condition, we applied the two-segment expression in a case study with a horizontally deployed source LSAA for spatial multiplexing region evaluation, see Table~\ref{table3}. Some interesting results associated with the 3D and 2D orientation constraints have been revealed. 
For instance, it is obvious that the optimal control of the receive array orientation in 3D is better than (or equally good as) optimal control in 2D. However, in the average sense, the DOF performance under uniform random orientation in 3D is actually worse than uniform random orientation in the 2D ground plane, and the spatial multiplexing region is larger in the later case.

\begin{appendices}

\section{Asymptotic Analysis for \texorpdfstring{$W_{\mathrm z}(R;\theta)$}{Wz}}
\label{Sec:Appendix1}

\subsection{Asymptotes derivation}
\label{Sec:A1_a}

Two functions $f(x)$ and $\tilde{f}(x)$ are said to be asymptotically equivalent as $x\rightarrow x_0$, if and only if $\lim_{x\rightarrow x_0} {f(x)}/{\tilde{f}(x)} =1$. This relation is denoted by $f(x) \sim \tilde{f}(x) \quad (x \rightarrow x_0)$. 
We aim to find functions asymptotically equivalent to $W_{\mathrm z}(R;\theta) \triangleq w_{\mathrm z}(0;\boldsymbol{\Omega})$ in different regimes of $R$, based on the expression of $w_{\mathrm z}(l;\boldsymbol{\Omega})$ given by \eqref{eq:wz0}. 
For convenience, we focus on $\theta \in (0,\frac{\pi}{2}]$ in the analysis such that $\sin\theta \geq 0$ and $\cos\theta \geq 0$, and extend the results to $\theta \in (\frac{\pi}{2},\pi)$ by symmetry. Moreover, we let $x = {R}/{L_s}$ and consider $f_{\mathrm z}(x;\theta) = \lambda \, W_{\mathrm z}(x L_s;\theta)$ instead. Based on \eqref{eq:wz0}, we obtain 
\begin{align}\label{eq:sw_z}
f_{\mathrm z}(x;\theta) = \frac{x \cos\theta + 0.5 }{g_1(x)} - \frac{x\cos\theta - 0.5 }{g_2(x)}, 
\end{align} 
where 
\begin{subequations}
\label{eq:g12}
\begin{align}
    g_1(x) = \sqrt{x^2 + x \cos\theta + 0.25},  
    \label{eq:g1} \\
    g_2(x) = \sqrt{x^2 - x \cos\theta + 0.25}. 
    \label{eq:g2} 
\end{align}
\end{subequations}
We treat $\theta$ as a given parameter and look for linear relationships between $\log( \tilde{f}_{\mathrm z})$ and $\log (x)$: 
\begin{align*}
\label{eq:asymptote_form}
    \log(\tilde{f}_{\mathrm z}(x)) = \log(A_{\mathrm z}) - B_{\mathrm z} \log (x), \quad \text{i.e.} \quad \tilde{f}_{\mathrm z}(x) = A_{\mathrm z} x^{-B_{\mathrm z}},
\end{align*}
where $B_{\mathrm z}$ is what we call the spatial bandwidth exponent (SBE). 

\subsubsection{Small \texorpdfstring{$R$}{r} regime} 
\label{Sec:A1_a1}

As $x \rightarrow 0^+$, both $x\cos\theta$ and $x \sin\theta$ become negligible compared to $\frac{L_s}{2}$. As a result, $f_{\mathrm z}(x) \sim 2 \  (x \rightarrow 0^+)$. Accordingly, we obtain an asymptote of $W_{\mathrm z}(R;\theta)$ for small $R$ values (with $B_{\mathrm z,1} =0$): 
\begin{equation}\label{eq:asym_z1}
	W_{\mathrm z}(R;\theta) \sim \tilde{W}_{\mathrm z}^{(1)}(R;\theta)  = \frac{2}{\lambda} \quad (R\rightarrow 0^+). 
\end{equation} 

\subsubsection{Large \texorpdfstring{$R$}{r} regime}
\label{Sec:A1_a2}

We rewrite \eqref{eq:sw_z} as 
\begin{align*}
f_\mathrm{z}(x;\theta) 
= \frac{0.5 \left[ g_1(x) + g_2(x) \right] - x\cos\theta \left[ g_1(x) - g_2(x) \right] }{ g_1(x) g_2(x) }, 
\end{align*}
where $g_1(x) - g_2(x)$ can be easily shown to be $ \frac{2 \cos\theta }{ 
 \sqrt{ 1 + \cos\theta x^{-1}  + 0.25 x^{-2} } + 
 \sqrt{ 1 - \cos\theta x^{-1}  + 0.25 x^{-2} } }$.  
As $x\rightarrow \infty$, $g_1(x) - g_2(x) \sim \cos\theta$ since the ``$1$'' term in the square-root operators dominates, whereas $g_1(x) + g_2(x) \sim 2x$ and $ g_1(x) g_2(x) \sim x^2$ since the ``$x^2$'' term in the square-root operators in their expressions given in \eqref{eq:g12} dominates. As a result, 
$f_{\mathrm z}(x;\theta) \sim \frac{x - x \cos^2\theta}{x^2} = \frac{\sin^2\theta}{x} \ (x\rightarrow \infty)$, which lead to an asymptote of $W_{\mathrm z}(R;\theta)$ for large $R$ values (with $B_{\mathrm z,3} =1$): 
\begin{align}\label{eq:asym_z3}
	W_{\mathrm z}(R;\theta) \sim \tilde{W}_{\mathrm z}^{(3)}(R;\theta) = \frac{\sin^2\theta}{\lambda} \cdot \frac{L_s}{R} \quad (R \rightarrow \infty). 
\end{align}

\subsubsection{Medium \texorpdfstring{$R$}{r} regime}
\label{Sec:A1_a3}

By observing \eqref{eq:sw_z} - \eqref{eq:g12}, we expect the tangent line of $\log(f_{\mathrm z}(x))$ at $\log(x = x_0)$, where $x_0 = \frac{1}{2 \cos\theta}$, to be a plausible asymptote candidate  for intermediate values of $x$. Note that we need to assume $\theta \neq \frac{\pi}{2}$ for the derivation and will run into trouble when $\theta \rightarrow \frac{\pi}{2}$, since then $x_0$ corresponds to a value in the large $R$ regime. Fortunately, as we will see in a moment, as $\theta\rightarrow \frac{\pi}{2}$, the SBE of the obtained asymptote, denoted by $B_{\mathrm z,2}$, approaches $1$, and the asymptote, denoted by $\tilde{W}_{\mathrm z}^{(2)}(R,\theta)$ approaches $\tilde{W}_{\mathrm z}^{(3)}(R;\frac{\pi}{2})$. This prevents large approximation errors when applying $\tilde{W}_{\mathrm z}^{(2)}(R,\theta)$ for $\theta$ values close to $\frac{\pi}{2}$.

The tangent line can be written explicitly as: 
\begin{align}
    \log (\tilde{f}_{\mathrm z}(x) ) &= \log ( f_{\mathrm z} (x_0 ) ) - B_{\mathrm z,2} [ \log(x) - \log (x_0 ) ] ,
    \label{eq:asymp_z_mid}
\end{align}
where
\begin{align}
    B_{\mathrm z,2} = - \frac{\mathrm d \log f_{\mathrm z}(x)  }{\mathrm d \log (x)} \Bigg|_{x = x_0} 
    = - \left[ \frac{1}{f_{\mathrm z}(x)} \cdot  \frac{\mathrm d f_{\mathrm z}(x)}{\mathrm d x} \cdot x  \right] \Bigg|_{x = x_0 } 
    \label{eq:alphaZ}
\end{align}
is $\theta$ dependent. Based on \eqref{eq:sw_z}, it can be directly obtained that 
\begin{align}\label{eq:fz_x0}
    f_{\mathrm z} (x_0) = \frac{1}{g_1(x_0)}= \sqrt{1- \eta^2(\theta) },
\end{align}
where $\eta (\theta) \triangleq \frac{\sin \theta}{ \sqrt{1 + 3 \cos^2 \theta } }$, 
 as given by \eqref{eq:eta_1}. Following the quotient rule, we derive
\begin{align}
\frac{\mathrm d  f_{\mathrm z}(x)}{\mathrm d x}
	= &\frac{\cos\theta }{ g_1(x) } - \frac{(x \cos\theta+0.5)(x+ 0.5 \cos\theta)}{ g_1(x)^3} \nonumber\\
 &- \frac{\cos\theta}{g_2(x)} + \frac{(x \cos\theta -0.5)(x- 0.5\cos\theta)}{ g_2(x)^3}. \label{eq:diff_f_to_t} 
\end{align}
Substituting \eqref{eq:fz_x0} and \eqref{eq:diff_f_to_t} into  \eqref{eq:alphaZ} and letting $x = x_0$, we obtain $B_{\mathrm z,2} (\theta) = \frac{ 1}{2} \left[  \eta^2(\theta) + {\eta^{-1} (\theta) } \right]$ as given in \eqref{eq:alpha}. 
Substituting $B_{\mathrm z,2} (\theta)$ and \eqref{eq:fz_x0} into \eqref{eq:asymp_z_mid}, and recalling $x = \frac{R}{L_s}$, we obtain the following asymptote: 
\begin{align}
\label{eq:asym_z2}
\tilde{W}_{\mathrm z}^{(2)}(R;\theta) = \frac{1}{\lambda}\sqrt{1- \eta^2(\theta) } \, \left(\frac{L_s}{2 \cos\theta \, R} \right)^{B_{\mathrm z,2}(\theta)}, 
\end{align}
such that $ W_{\mathrm z}(R;\theta) \sim \tilde{W}_{\mathrm z}^{(2)}(R; \theta) \quad (R\rightarrow \frac{L_s}{2\cos\theta} ) $. 

It is easy to see that $\eta(\theta)$ is an increasing function over $\theta \in (0,\frac{\pi}{2})$ and $\eta(\theta)\sim1 \  (\theta \rightarrow \frac{\pi}{2})$. Hence,  $B_{\mathrm z,2} (\theta)\sim 1 \  (\theta \rightarrow \frac{\pi}{2})$. With this and $\sqrt{1- \eta^2(\theta) } = \frac{2\cos\theta}{\sqrt{1 + 3 \cos^2 \theta }}$, it can be verfied that $\tilde{W}_{\mathrm z}^{(2)}(R; \theta) \sim \tilde{W}_{\mathrm z}^{(3)}(R; \theta) \ (\theta \rightarrow \frac{\pi}{2})$.

\begin{figure}[!t]
	\centering 
	\includegraphics [width= \linewidth,trim= 0 0 0 0, clip]{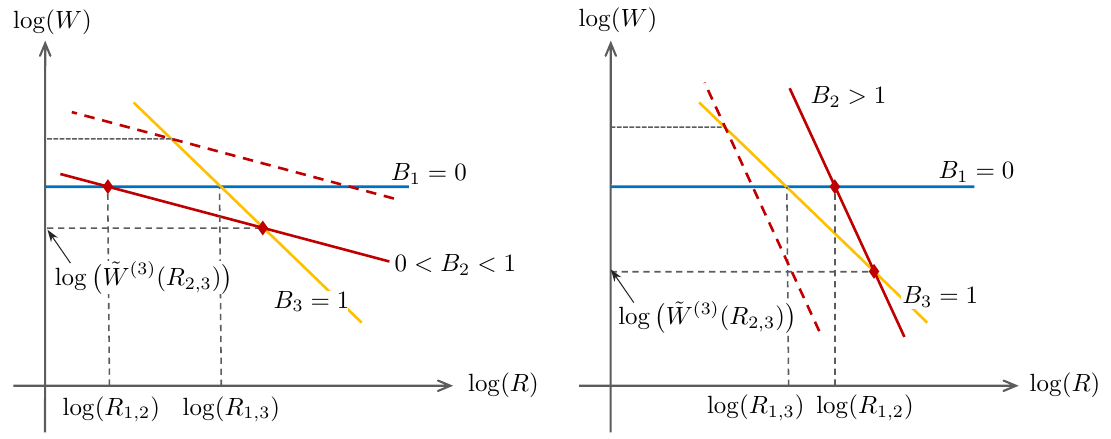} 
	\caption{Possible geometric relationships of three asymptotes $\Tilde{W}^{(k)}$, $\log(\Tilde{W}^{(k)}) \propto (\log(R) )^{-B_k}$, for $k=1,2,3$. The dashed lines represent the situations where $\Tilde{W}^{(2)}$ is not applicable.}  
 \vspace{-1em}
\label{fig:asymp_cd}
\end{figure}

\subsection{Applicability discussion}
\label{Sec:A1_b}

Whether two or three of the derived asymptote segments, $\tilde{W}_{\mathrm z}^{(1)}(R; \theta)$, $\tilde{W}_{\mathrm z}^{(2)}(R; \theta)$, and $\tilde{W}_{\mathrm z}^{(3)}(R; \theta)$, should be used to form the asymptotic function depends on how they intersect. It can be easily seen that 
$\tilde{W}_{\mathrm z}^{(1)}(R;\theta)$ always intersects $\tilde{W}_{\mathrm z}^{(2)}(R;\theta)$ and $\tilde{W}_{\mathrm z}^{(3)}(R;\theta)$. However, when $\theta$ approaches $\theta_{\mathrm z,1}$ or $\frac{\pi}{2}$, $B_{\mathrm z,2}(\theta) \to 1$, which implies that $\tilde{W}_{\mathrm z}^{(2)}(R;\theta)$ and  $\tilde{W}_{\mathrm z}^{(3)}(R;\theta)$ are asymptotically parallel.  We therefore exclude $\tilde{W}_{\mathrm z}^{(2)}(R;\theta)$ from the asymptotic function. Intersection between $\tilde{W}_{\mathrm z}^{(2)}(R;\theta)$ and  $\tilde{W}_{\mathrm z}^{(3)}(R;\theta)$ is guaranteed under any other $\theta$ conditions, but $\tilde{W}_{\mathrm z}^{(2)}(R;\theta)$ should only be included if the intersection is below $\tilde{W}_{\mathrm z}^{(1)}$.

By drawing the possible locations of the three asymptotes in the $\log(W)$-$\log(R)$ plane, which is shown in Fig.~\ref{fig:asymp_cd}, we can see that this requirement translates to the two conditions given by \eqref{eq:condition_z1} and \eqref{eq:condition_z2}, quoted below: 
\begin{align*}
 0<B_{\mathrm z,2} < 1, \  R_{\mathrm z,1,2}(\theta) < R_{\mathrm z,1,3}(\theta),  \nonumber\\ 
  B_{\mathrm z,2} > 1, \ R_{\mathrm z,1,2}(\theta) > R_{\mathrm z,1,3}(\theta).     
\end{align*}

The critical distances where any two asymptotes intersect can be easily found. Since $\tilde{W}_{\mathrm z}^{(1)}(R;\theta)$,  $\tilde{W}_{\mathrm z}^{(2)}(R;\theta)$, and $\tilde{W}_{\mathrm z}^{(3)}(R;\theta)$ can all be expressed in the form of $A R^{-B}$, and the solution of equation $A_1 R^{-B_1} = A_2 R^{-B_2}$ is given by $R = (\frac{A_1}{A_2})^{\frac{ 1 }{B_1 - B_2 }}$, the critical distances $R_{\mathrm z,1,2}(\theta)$, $R_{\mathrm z,2,3}(\theta)$, and $R_{\mathrm z,1,2}(\theta)$, given in Table~\ref{table1}, can be easily derived.

From the expression for $B_{\mathrm z,2} (\theta)$, we see that to solve $B_{\mathrm z,2} (\theta_{\mathrm z,1}) =1$ we need first to solve the equation $ x^2 + x^{-1} =2$. Two positive roots are obtained: $x_1=\frac{\sqrt{5} -1}{2}$ and $x_2=1$. Letting $\eta(\theta_{\mathrm z,1}) = x_1$, the expression $ \theta_{\mathrm z,1} = \arccos \left( \sqrt{ \frac{1}{2\sqrt{5} -1} } \,\right) \approx 0.3197\pi$ is obtained. When $\theta < \theta_{\mathrm z,1}$, $B_{\mathrm z,2} >1$; and when $\theta> \theta_{\mathrm z,1}$, $0< B_{\mathrm z,2} <1$. Moreover, as shown in Fig.~\ref{fig_parametersZ}(b), for $\theta \in (0, \theta_{\mathrm z,2})$, $R_{\mathrm z,1,3}(\theta)< R_{\mathrm z,1,2}(\theta)$; and for $\theta \in (\theta_{\mathrm z,2}, \frac{\pi}{2})$, $R_{\mathrm z,1,3}(\theta)> R_{\mathrm z,1,2}(\theta)$, where $\theta_{\mathrm z,2} \approx 0.3285\pi > \theta_{\mathrm z,1}$. As a result, for $\theta \in [\theta_{\mathrm z,1}, \theta_{\mathrm z,2}]$, asymptote $\tilde{W}_{\mathrm z}^{(2)}(R;\theta)$ should not be used  in the asymptotic function.

Since $W_{\mathrm z}(R;\theta)$ is symmetric about $\theta = \frac{\pi}{2}$, by replacing $\cos\theta$ using $|\cos\theta|$ in \eqref{eq:asym_z3} and \eqref{eq:asym_z2}, the applicability of the obtained asymptote functions is also extended to the range $\theta \in (\frac{\pi}{2},\pi)$. 
Having all the above discussions, the formation rule of multi-slope asymptotic function for $W_{\mathrm z}(R;\theta)$, summarized in the last paragraph of Section~\ref{sec:3a} and also in Table~\ref{table1}, can be concluded.

\section{Asymptotic Analysis for \texorpdfstring{${W}_{\mathrm x}(R;\theta)$}{Wx}}
\label{Sec:Appendix2}

\subsection{Asymptotes derivation}

We first derive functions asymptotically equivalent to $W_{\mathrm x}(R;\theta) \triangleq w_{\mathrm x}(0;\boldsymbol{\Omega})$ in different regimes of $R$, based on the exact expression of $w_{\mathrm x}(l;\boldsymbol{\Omega})$ given by \eqref{eq:wx0}. For convenience, we again focus on $\theta \in (0,\frac{\pi}{2}]$, define $x = \frac{R}{L_s}$, and discuss $f_{\mathrm x}(x;\theta) = \lambda \, W_{\mathrm x}(x L_s;\theta)$ instead. Based on \eqref{eq:wx0}, we obtain 
\begin{align}\label{eq:sw_x}
f_{\mathrm x}(x;\theta) 
 &=  \begin{cases} 
	f_{\mathrm x 1}(x;\theta) = 1- \frac{x\sin\theta }{g_1(x)} , & x \leq \frac{1}{2 \cos\theta} \\
	f_{\mathrm x 2}(x;\theta) = \frac{x \sin\theta }{g_2(x)}  - \frac{x \sin\theta }{g_1(x)},  & x \geq \frac{1}{2 \cos\theta}
	\end{cases}  
\end{align} 
where $g_1(x)$ and $g_2(x)$ are given in 
\eqref{eq:g12}. Treating $\theta$ as a given parameter, we look for linear relationships between $\log( \tilde{f}_{\mathrm x}(x))$  and $\log (x)$ of the form 
\begin{align*}
    \log(\tilde{f}_{\mathrm x}(x)) = \log(A_{\mathrm x}) - B_{\mathrm x} \log (x), \quad \text{or} \quad \tilde{f}_{\mathrm x}(x) = A_{\mathrm x} x^{-B_{\mathrm x}}.
\end{align*}
Note that when $\theta = \frac{\pi}{2}$, $f_{\mathrm x 2}(x;\theta)$ in \eqref{eq:sw_x} will never apply since $\cos\theta = 0$. Therefore, special attention is required for this particular $\theta$ value. 

\subsubsection{Small \texorpdfstring{$R$}{r} regime} 
As $x \rightarrow 0^+$, it can be easily see that for any $\theta \in (0,\frac{\pi}{2}]$, $f_{\mathrm x}(x;\theta) = f_{\mathrm x 1}(x;\theta) \sim 1$. Accordingly, we obtain an asymptote of $W_{\mathrm x}(R;\theta)$ for small $R$ values (with $B_{\mathrm x,1} =0$): 
\begin{equation}\label{eq:asym_x1}
W_{\mathrm x}(R;\theta) \sim\tilde{W}_{\mathrm x}^{(1)}(R;\theta)  = \frac{1}{\lambda} \quad (R \rightarrow 0^+). 
\end{equation}  

\subsubsection{Large \texorpdfstring{$R$}{r} regime}   
As $x \rightarrow \infty$, $f_{\mathrm x}(x;\theta) = f_{\mathrm x 2}(x;\theta)$ if $\theta \in (0,\frac{\pi}{2})$. 
Recalling that $g_1(x) + g_2(x) \sim 2x$, $ g_1(x) g_2(x) \sim x^2$, and $g_1(x) - g_2(x) \sim  \cos\theta$ as $x \rightarrow \infty$, we have 
\begin{align*}
f_{\mathrm x 2}(x;\theta) = \frac{x \sin\theta \left[ g_1(x) - g_2(x)  \right]  }{ g_1(x) g_2(x)} \sim \frac{\sin\theta \cos\theta   }{x} \quad (x \rightarrow \infty). 
\end{align*}
Accordingly, we obtain an asymptote of $W_{\mathrm x}(R;\theta)$ for large $R$ values for $\theta \in (0,\frac{\pi}{2})$ (with $B_{\mathrm x,3} =1$): 
\begin{align}\label{eq:asym_x3}
W_{\mathrm x}(R;\theta) \sim \tilde{W}_{\mathrm x}^{(3)}(R;\theta) = \frac {\sin\theta \cos\theta }{\lambda} \cdot \frac{L_s}{R} \quad (R \rightarrow \infty). 
\end{align}

When $\theta = \frac{\pi}{2}$, no matter how large $x$ is, we always have $f_{\mathrm x}(x;\frac{\pi}{2}) = f_{\mathrm x 1}(x;\frac{\pi}{2})$, which can be written as 
\begin{align*}
f_{\mathrm x 1}(x;\frac{\pi}{2}) = 
    \frac{0.25}{x^2 \sqrt{1 + 0.25 x^{-2} } ( \sqrt{1 + 0.25 x^{-2} } + 1 ) }. 
\end{align*}
As $x\rightarrow \infty$, $\sqrt{1 + 0.25 x^{-2}} \rightarrow 1$, and thus $f_{\mathrm x1}(x;\frac{\pi}{2}) \sim \frac{1}{8x^2}$.
Accordingly, we obtain an asymptote of $W_{\mathrm x}(R;\theta)$ for large $R$ values for $\theta =\frac{\pi}{2}$ (with $B_{\mathrm x,3^*} =2$): 
\begin{align}\label{eq:asym_x3_2}
W_{\mathrm x}(R;\theta) \sim \tilde{W}_{\mathrm x}^{(3^*)}(R;\theta) =  \frac{1}{8\lambda} \cdot \left(\frac{L_s}{R}\right)^2 \quad (R \rightarrow \infty). 
\end{align}

\subsubsection{Medium \texorpdfstring{$R$}{R} regime}

Observing \eqref{eq:sw_x}, we expect the tangent lines of $\log(f_{\mathrm x 1}(x))$ or $\log(f_{\mathrm x 2}(x))$ at $\log(x =x_0)$, where $x_0 = \frac{1}{2 \cos\theta}$, to be plausible candidates of the asymptote. We again assume $\theta\neq \frac{\pi}{2}$ first, and after derivation following the same steps as for the $\mathrm z$ direction, it is found that $\log(f_{\mathrm x 1}(x))$ and $\log(f_{\mathrm x 2}(x))$ share the same tangent line at $\log(x_0)$. Below we will only provide the derivation regarding $\log(f_{\mathrm x 1}(x))$ in detail. Specifically, we aim to obtain
\begin{align}\label{eq:asymp_x_mid}
    \log (\tilde{f}_{\mathrm x}(x) ) &= \log ( f_{\mathrm x1} (x_0 ) ) - B_{\mathrm x,2} [ \log(x) - \log (x_0 ) ] ,
\end{align}
where 
\begin{align}
 B_{\mathrm x,2} = -\frac{\mathrm d \log f_{\mathrm x1}(x)  }{\mathrm d \log (x)} \Bigg|_{x = x_0}
    = -\left[ \frac{1}{f_{\mathrm x1}(x)} \cdot  \frac{\mathrm d f_{\mathrm x1}(x)}{\mathrm d x} \cdot x  \right] \Bigg|_{x = x_0 }
    \label{eq:betaX}
\end{align}
has dependency in $\theta$. It is easily computed that 
\begin{align} \label{eq:fx_value}
 f_{\mathrm x 1}(x_0)  = 1 -  \frac{\sin \theta}{ \sqrt{1 + 3 \cos^2 \theta } }  =  1-\eta(\theta),
\end{align} 
where $\eta(\theta)$ is given by  \eqref{eq:eta_1}. Following the quotient rule, 
\begin{align}\label{eq:fx1_derivative}
\frac{\mathrm d  f_{\mathrm x1}(x)  }{\mathrm d x}
= &  -\frac{\sin\theta }{ g_1(x) } + \frac{x \sin\theta (x+0.5 \cos\theta)}{g_1(x)^3}.
\end{align}
Substituting \eqref{eq:fx_value} and \eqref{eq:fx1_derivative} into \eqref{eq:betaX} and letting $x=x_0$, we obtain $B_{\mathrm x,2}(\theta) =\frac{1}{2} \big[ \eta^2(\theta)+ \eta(\theta) \big]$ 
as given in \eqref{eq:beta}. Substituting  $B_{\mathrm x,2}(\theta)$ and \eqref{eq:fx_value} into \eqref{eq:asymp_x_mid}, and recalling $x = \frac{R}{L_s}$, we obtain the following asymptote: 
\begin{align}
\label{eq:asym_x2}
\tilde{W}_{\mathrm x}^{(2)}(R,\theta) 
	&=  \frac{1}{\lambda} \big[1 - {\eta(\theta)} \big] \cdot \left(\frac{L_s}{2  \cos\theta \, R} \right)^{B_{\mathrm x,2}(\theta)} , 
\end{align}
such that  $W_{\mathrm x}(R,\theta) \sim \tilde{W}_{\mathrm x}^{(2)}(R,\theta) \quad (R\rightarrow \frac{L_s}{2\cos\theta})$. 

As noted already, when deriving $\tilde{W}_{\mathrm z}^{(2)}(R,\theta)$, $x_0$ corresponds to a value in the large $R$ regime when $\theta \rightarrow \frac{\pi}{2}$. We do not have the same luck as for the $\mathrm z$ direction. By applying $\tilde{W}_{\mathrm x}^{(2)}(R,\theta)$ for $\theta\rightarrow \frac{\pi}{2}$, large approximation errors will occur because $B_{\mathrm x,2} (\theta)$ no longer captures the correct decay rate in the actual medium $R$ regime. As we have already derived, $f_{\mathrm x 1}(x;\frac{\pi}{2})$ decays with SBE $B_{\mathrm x,3^*} =2$ for large $x$. In fact, by plotting the curve of $f_{\mathrm x 1}(x;\theta)$ for some $\theta$ value very close to $\frac{\pi}{2}$, one can observe that before decaying with exponent $B_{\mathrm x,2}(\theta) \approx 1$ at $x = x_0$, $f_{\mathrm x 1}(x;\theta)$ first decays with an exponent $2$ for some smaller $R$ values; and as $R$ increases, the exponent gradually evolves to $B_{\mathrm x,2}(\theta)$. Nevertheless, at the same $R$ in the medium regime, the spatial bandwidth admitted under these $\theta$ conditions is much smaller than under those favorable conditions, that is, when $\theta \simeq \frac{\pi}{4}$. Therefore, we do not aim for more accurate asymptotes for medium $R$ for $\theta \simeq \frac{\pi}{2}$.

\subsection{Applicability discussion}

Asymptotic functions can now be formed using the four derived asymptotes. For $\theta = \frac{\pi}{2}$, $\tilde{W}_{\mathrm x}^{(1)}(R,\theta)$ and $\tilde{W}_{\mathrm x}^{(3^*)}(R,\theta)$ are used to form a dual-slope asymptotic function, and their applicable ranges are separated by critical distance $R_{\mathrm x,1,3^*}(\theta) = \frac{L_s}{\sqrt{8}}$, at which the two asymptotes intersect. For $\theta \in (0, \frac{\pi}{2})$, two or three of $\tilde{W}_{\mathrm x}^{(1)}(R,\theta)$, $\tilde{W}_{\mathrm x}^{(2)}(R,\theta)$, and $\tilde{W}_{\mathrm x}^{(3)}(R,\theta)$ are used, depending on how they intersect. Since $0<B_{\mathrm x,2}(\theta)<1$, the segments always intersect, and the critical distances of the intersections:  $R_{\mathrm x,1,2}(\theta)$, $R_{\mathrm x,2,3}(\theta)$, and $R_{\mathrm x,1,2}(\theta)$ given in Table~\ref{table2}, can be easily derived. However, $\tilde{W}_{\mathrm x}^{(2)}(R,\theta)$ is used only when $R_{\mathrm x,1,2}(\theta) < R_{\mathrm x,1,3}(\theta)$. As shown in Fig. \ref{fig_parametersX}~(b), this condition is violated when $\theta \leq \theta_{\mathrm x}$, where $\theta_{\mathrm x} \approx 0.0225\pi$ is the numerical solution to $R_{\mathrm x,1,2}(\theta) = R_{\mathrm x,1,3}(\theta)$. Due to the symmetry of $W_{\mathrm x}(R;\theta)$ with respect to $\theta = \frac{\pi}{2}$, by replacing $\cos\theta$ with $|\cos\theta|$ in \eqref{eq:asym_x3} and \eqref{eq:asym_x2}, the applicability of the obtained asymptote functions is extended to the range $\theta \in (\frac{\pi}{2},\pi)$. With the above discussion in mind, the formation rule of multi-slope asymptotic function for $W_{\mathrm x}(R;\theta)$, summarized in the last paragraph of Section~\ref{sec:3b} and also in Table~\ref{table2}, can be concluded.  

\end{appendices}

\bibliographystyle{IEEEtran}
\bibliography{ref}

\end{document}

%% file: acronyms.tex
\begin{acronym}[ACRONYM]

\acro{EM}{electromagnetic}
\acro{AI}{artificial intelligence}
\acro{BP}{belief propagation}
\acro{KPI}{key performance indicator}
\acro{QoS}{Quality of Service}
\acro{RF}{radio frequency}
\acro{UWB}{ultra wideband}

\acro{Tx}{transmitter}
\acro{Rx}{receiver}
\acro{TXRX}[TX/RX]{transmitter/receiver}
\acro{BS}{base station}
\acro{UE}{user equipment}
\acro{SNR}{signal-to-noise ratio}
\acro{UL}{uplink}
\acro{DL}{downlink}
\acro{IQ}{in-phase and quadrature}
\acro{RAN}{radio access network}
\acro{RAT}{radio access technology}

\acro{CDF}{cumulative density function}
\acro{PMF}{probability mass function}
\acro{PDF}{probability density function}
\acro{CCDF}{complementary cumulative distribution function}

\acro{RMSE}{root mean squared error}
\acro{LS}{least-squares}
\acro{ML}{maximum likelihood}
\acro{FIM}{Fisher information matrix}
\acro{CRB}{Cram\'er-Rao bound}
\acro{MCRB}{misspecified Cram\'er-Rao bound}

\acro{GDOP}{geometric dilution of precision}
\acro{GNSS}{global navigation satellite system}
\acro{GPS}{global positioning system}
\acro{IMU}{inertial measurement unit}
\acro{RTK}{real-time kinematic}
\acro{PRS}{positioning reference signal}

\acro{AoA}{angle-of-arrival}
\acro{AoD}{angle-of-departure}
\acro{ToA}{time-of-arrival}
\acro{TDoA}{time-difference-of-arrival}
\acro{RTT}{round-trip-time}

\acro{LOS}{line-of-sight}
\acro{NLOS}{non-line-of-sight}

\acro{1D}{one-dimensional}
\acro{2D}{two-dimensional}
\acro{3D}{three-dimensional}
\acro{DOF}{degree of freedom}
\acrodefplural{DOF}{degrees of freedom}

\acro{LCS}{local coordinate system}
\acro{GCS}{global coordinate system}

\acro{OFDM}{orthogonal frequency-division multiplexing}
\acro{mmWave}{millimeter-wave}
\acro{MIMO}{multiple-input multiple-output}
\acro{XL-MIMO}{extra-large/extremely large-scale MIMO}
\acro{D-MIMO}{distributed multiple-input multiple-output}
\acro{RIS}{reconfigurable intelligent surface}
\acro{XL-MIMO}{extremely large-scale multiple-input-multiple-output}
\acro{ELAA}{extremely large aperture array}
\acro{XL-array}{extremely large-scale/ array}
\acro{LSAA}{large-scale antenna array}
\acro{LIS}{large intelligent surface}

\acro{SVD}{singular value decomposition} 

\end{acronym}